\documentclass[
    journal,
    twocolumn,
    ]{IEEEtranTCOM}
\usepackage{graphicx}
\usepackage{amsmath}
\usepackage{amssymb}
\usepackage{amsthm}
\usepackage{afterpage}
\usepackage{verbatim}
\usepackage{psfrag}
\usepackage{color}
\usepackage{cite}
\usepackage{setspace}
\usepackage{epsfig}
\usepackage{epstopdf}
\usepackage{footmisc}
\usepackage{fmtcount}
\usepackage{stackrel}
\usepackage{float}
\usepackage{breqn}
\usepackage{enumerate}
\usepackage{graphicx}
\usepackage{subcaption}
\usepackage{graphicx}

\begin{document}
\title{Towards the Internet of Underground Things: A Systematic Survey}
\author{Nasir Saeed,~\IEEEmembership{Senior Member,~IEEE}, Mohamed-Slim Alouini,~\IEEEmembership{Fellow,~IEEE}, Tareq Y. Al-Naffouri,~\IEEEmembership{Senior Member,~IEEE}
\thanks{This work is
supported by Office of Sponsored Research (OSR) at King Abdullah University of Science and Technology (KAUST). 

The authors are with the Computer Electrical and Mathematical Sciences \& Engineering (CEMSE) Division, KAUST, Thuwal, Makkah Province, Kingdom of Saudi Arabia, 23955-6900.}
}
\maketitle
\begin{abstract}
This paper provides recent advances in the area of Internet of Underground Things (IoUT) with emphasis on enabling communication technologies, networking issues, and localization techniques. IoUT is enabled by underground things (sensors), communication technology, and networking protocols. This new paradigm of IoUT facilitates the integration of sensing and communication in the underground environment for various industries, such as oil and gas, agriculture, seismic mapping, and border monitoring. These applications require to gather relevant information from the deployed underground things. However, the harsh underground propagation environment including sand, rock, and watersheds do not allow the use of single communication technology for information transfer between the surface and the underground things. Therefore, various wireless and wired communication technologies are used for underground communication. The wireless technologies are based on acoustic waves, electromagnetic waves, magnetic induction and visible light communication while the wired technologies use coaxial cable and optical fibers. In this paper, state-of-art communication technologies are surveyed, and the respective networking and localization techniques for IoUT are presented. Moreover, the advances and applications of IoUT are also reported. Also, new research challenges for the design and implementation of IoUT are identified.
\end{abstract}
\IEEEpeerreviewmaketitle

\section{Introduction}
The population of the world will increase by 31 \% in 2050  \cite{prb}, and therefore will require more natural resources and food to survive. In the next three decades with such increase in population, 71 \% more resources are required {    \cite{ung}}. This ever-increasing demand for resources needs novel technologies to improve the underground exploration for natural resources and to produce more crop. The subsurface environment and agricultural lands provide various natural resources, such as earth minerals, fossil fuels, metal ores, groundwater, and food. To efficiently use all of these resources, Internet of Underground Things (IoUT) is an enabling technology which can provide smart oil and gas fields, smart agriculture fields, and smart seismic quality control. However, implementation of IoUT is a challenging task due to the harsh underground environment which requires low power and small size underground sensors, long-range communication technology, efficient networking solutions, and accurate localization techniques.

{

The above applications and challenges of IoUT lead to active research in this area. The significant difference between the in-air internet of things and IoUT is the communication media where the sensors (underground things) are buried and communicate through the soil. {     Note that there is also a possibility to place the underground things in an open underground space, such as mines and tunnels. In this case, the network setup is underground, but the communication between the devices takes place through the air and hence terrestrial wireless technologies, such as radio frequency (RF) and visible light (VLC) are used. There are existing review articles on information and communication systems for mines and tunnels. 
For example, in \cite{Singh2018}, the authors briefly discuss the information and communications systems for mining IoT. Similarly, the authors in \cite{Mudul2018} presented a survey on the applications of wireless sensor networks for underground coal and mines. However, communication through the soil is different where the transmission signal suffers from various underground impairments. Hence, in this paper, we collect the literature on communication, networking, and localization for buried smart objects. {\footnote{    IoT in mining is beyond the scope of this survey since there are existing review articles on this subject, such as \cite{Singh2018, Mudul2018, Zhou2017}.}}}

Due to the heterogeneous nature of the soil which consists of sand, rock, and watersheds, communication through it is more challenging. In the past, various solutions were used for underground communications. For example, the use of mud pulse telemetry (MPT) communication systems for oil and gas monitoring dates back to the mid of the $\text{19}^{\text{th}}$ century \cite{Arps1964}. The MPT systems work on the concept of mud circulation in the pipes for data transmission \cite{thakur2018most}. Although MPT systems are well-developed for down-hole monitoring, their data rate is low, i.e., in bits per second. To improve the data rates for underground communications, wired cables, such as coaxial cables and fiber optics are also used. These wired solutions provide high data rates, timely, reliable and accurate solutions, especially for the deep underground monitoring. Therefore, wired technologies including coaxial cable and optical fiber are used in many works, such as \cite{Schroeder2002, hernandez2008high,algeroy2010permanent} for down-hole monitoring. Although wired solutions have the above advantages, they have high complexity and are not scalable solutions. Therefore, wireless solutions are investigated to provide low complex, high data rate, and scalable solutions. 

\begin{table*}[htb!]
\footnotesize
\centering
\caption{{    Comparison  of underground communication technologies for IoUT.}}
\label{Tableuwcsystems}
\begin{tabular}{| p{1.7cm} | p{2.4cm} | p{2.1cm} | p{2.1cm}| p{2.1cm} | p{2.1cm}| p{2.1cm}|}
\hline
\hline
 \textbf{Parameters}              & \textbf{EM}  & \textbf{Acoustic}    & \textbf{Mud pulse} & \textbf{MI}&  \textbf{Wired}& \textbf{VLC} \\ \hline
 \textbf{Transmission Range}   & few meters             & In hundred of meters  & In hundred of meters  &In tens of meters & In hundred of meters  & In tens of meters     \\ \hline
 \textbf{Attenuation}            & High  & High & Medium & Low & Low & High \\ \hline
 \textbf{Interference}            &High & Medium & Medium & Low & Low & Low   \\ \hline
 \textbf{Installation cost}          & Medium & Medium & Low & Medium & Low & Low  \\ \hline
 \textbf{Data rate}               &  In tens of bps      &  	In tens of bps  &  	In tens of bps        & In {   kbps} & In Mbps & In {   kbps}\\ \hline
 \textbf{Applications}               &  Agriculture, seismic exploration, and down-hole telemetry     &  	Seismic exploration, buried pipeline monitoring, and down-hole telemetry    &  	Down-hole telemetry        & Down-hole telemetry & Down-hole telemetry and buried pipeline monitoring & Down-hole telemetry\\ \hline

{     \textbf{Advantages}  }             &  {    High data rate compare to acoustic and MPT systems, easy to install, and can be implemented in multi-hop fashion}     &  	{    Long transmission range and high data rate than MPT systems}   &  	{    Long communication range and low installation cost}        & {    High data rate, support multi-hop communication, low interference, and low attenuation} & {    High data rate, reliable, low latency, and accurate}  & {    Low installation cost and low interference}\\ \hline 
 {    \textbf{Disadvantages}  }             &  {    High attenuation and interference, limited transmission range, and require large antennas}    & {    High attenuation and low data rate than EM, MI, wired, and VLC solutions} 	 &  {    Low data rate and high complexity}  & {    Limited transmission range, need dense deployment, and orientation between the coils is required}    & {    High complexity and non-scalable} & {    High attenuation, require line of sight, and can operate only if the underground medium is air}\\ \hline
\hline
\hline       
\end{tabular}
\end{table*}

These wireless solutions include acoustic waves, electromagnetic waves (EM), magnetic-induction (MI), and VLC. Acoustic waves are used for the detection of objects underground, soil moisture detection, and down-hole communications. For example, in \cite{Oelze2002}, the authors used 2 to 6 {   kHz} frequencies to detect an underground object. Similarly, a frequency of 900 {   Hz} was used in \cite{sharma2010continuous} to find the soil moisture. Recently, the authors in \cite{Singer2017} and \cite{Sijung2018} proposed acoustic waves based wireless data transmission system (SoilComm) for IoUTs that was able to transmit the sensing data over 30 m distance through the soil. Acoustic waves-based solutions are good for detection purposes, such as underground object detection and soil moisture detection. However, they provide low data rate communication, i.e., in tens of bits per second for down-hole monitoring and also suffers from the acoustic noise and attenuation of the acoustic signal along the drill-pipe \cite{gardner2006acoustic}.

To investigate the use of EM waves for underground communications, various frequencies of the EM spectrum were examined in the past starting from below 500 {   kHz} \cite{Yoon2012, Yoon2012a, Ghazanfari2012} to 120 THz \cite{Akkas2012}. Depending on the frequency of EM waves, the transmission range of EM signal varies from a few meters (few hundred {   kHz}) to centimeters (THz). Due to their low penetration depth in the soil, EM waves are mainly used for agriculture applications. For example, in \cite{VURAN2018} the authors presented EM waves-based IoUT solution for precision agriculture. The EM waves suffer from severe path loss in soil, and therefore their transmission range is low. Especially underground watersheds have a notoriously bad impact on the propagation of EM waves by limiting its transmission range. Alternatively, multi-hop MI-based underground communications is examined in the recent past \cite{Sun2009}.
Although the MI channel is more robust to the underground environment, it suffers from low transmission range and requires perfect orientation of the transmitter and receiver coils which might be challenging in the underground environment. The research on developing MI-based IoUT is still in the academic phase and faces various challenges.
Nevertheless, VLC is also being investigated for the underground communications in gas reservoirs \cite{Miramirkhani2018}. However, in the VLC based system, the propagation of light is affected by the gas as well as it requires perfect alignment between the LEDs and the photo-detector. In short, among all these solutions, MPT systems, coaxial cable, fiber optics are the commercially available technologies for down-hole monitoring while acoustic, EM, MI, and VLC based systems are still in the academic phase of research and are only tested by lab experiments. Table \ref{Tableuwcsystems} compares various communication technologies for IoUT.   

}

\begin{figure*}[t]
  \centering
  \includegraphics[width=1\textwidth]{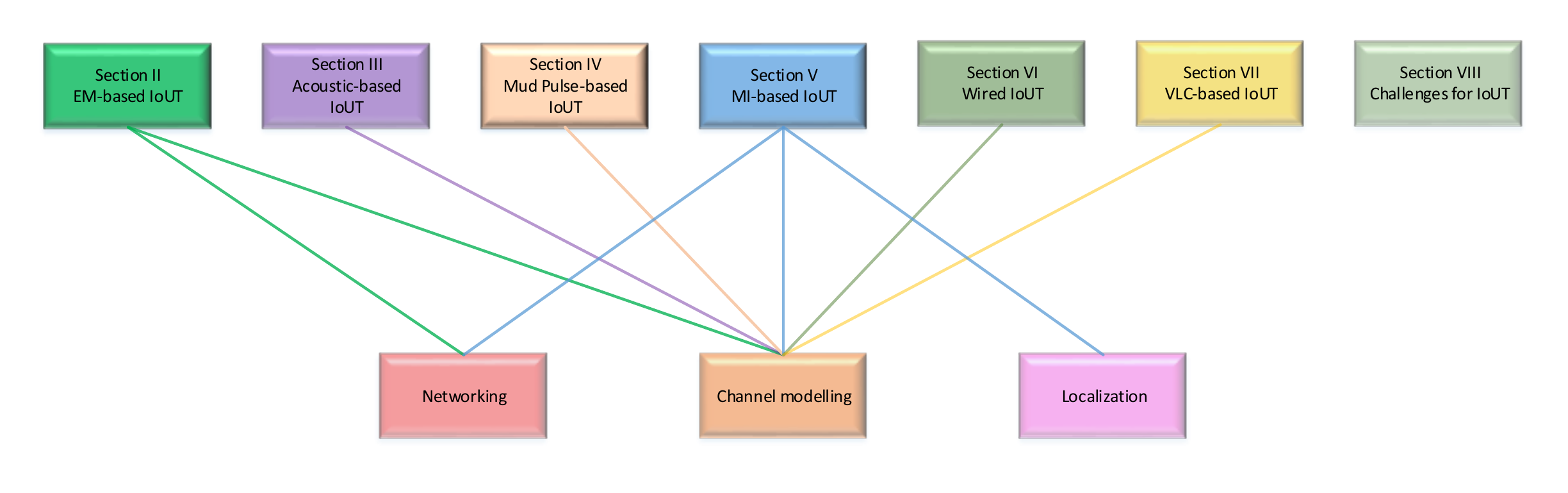}\\
  \caption{Organization of the survey.}
  \label{OCC_Organization}
\end{figure*}
\subsection{Related Surveys}
There are quite few survey articles published that cover various issues of IoUTs. For example, the work in \cite{VURAN2018} presents the EM waves based IoUTs for precise agriculture.  Moreover, \cite{VURAN2018} also reviews the academic testbeds and commercial solutions for precise agriculture. In \cite{Kisseleff2018}, the authors give an overview of MI-based underground wireless sensor networks and present challenges and applications. {   The authors in \cite{AALSALEM2018} present} the recent advances and challenges for wireless sensor networks in the oil and gas industry. 

The contributions of this article relative to the existing literature on IoUT is summarized as follows:
\begin{itemize}
\item Compared to existing papers for IoUTs, this paper provides a deeper understanding of the all relevant communication technologies, networking solutions, and localization techniques which can be used to implement various IoUT-based applications.
\item The existing surveys only {   present} the literature on EM and MI-based underground wireless communication networks. However, we also collect the research on acoustic, mud-pulse telemetry, visible light, and wired-based communication technologies for various IoUT applications.
\item We survey the key challenges to implement IoUT and explore the relationship between IoUT, big data analytics, cloud, and fog computing.
\end{itemize}

\subsection{Survey Organization}
The main focus of the paper is to review channel modeling, networking, and localization methods for the current sensing and communication technologies used in IoUT. However, for some of the techniques, such as acoustic, visible light communications, and mud-pulse telemetry, the networking, and localization problem is still an open research problem. {     Fig. \ref{OCC_Organization} illustrates the organization of this survey. In Section II, we present the literature on channel modeling and networking for EM-based underground communication. Sections III and IV cover the channel model for acoustic and mud pulse telemetry-based IoUTs, respectively. In Section V, channel modeling, networking, and localization for MI-based IoUTs are presented. VLC and wired based solutions are presented in section VI and VII, respectively. Section VIII discusses the advances and challenges of IoUTs in terms of each communication technology. Finally, section IX summarizes and concludes the survey.
}


\section{EM Waves for IoUT}
{    
EM waves are widely used for underground communication and sensing to enable various applications, such as smart agriculture \cite{DONG2013,Yu2017, Liedmann2017,Liedmann2018}, smart seismic exploration \cite{Savazzi2013}, smart drilling, and smart oil and gas fields \cite{Franconi2014, Akkas2016, Karli2017}.  In this section, we cover the literature on the channel modeling and networking for EM-based IoUTs.}
 
\subsection{Channel Modeling}
Channel modelling for EM-based underground communications dates back to the early 70's of the $\text{20}^{\text{th}}$ century. James \textit{et. al} investigated the propagation of EM signals through the earth surface where the frequency range of 1 to 10 MHz was experimentally tested in different types of soil \cite{Wait1971}. {    Their results suggest that at low frequencies of the EM spectrum, earth refractive index is large which simplifies the analysis. However, low frequencies suffer from limited bandwidth and low resolution for time of arrival pulses. Therefore, higher frequencies of the EM spectrum were tested in \cite{Wait1971} which suffers from high attenuation but have a low dispersion.} The authors in \cite{Sivaprasad1973} further examined the impact of different layers of the soil on the propagation of EM signals where Fast Fourier Transform was used to find the reflection of the incident signals from a three-layered medium. Furthermore, Lytle \textit{et. al} measured the electrical characteristics of the earth medium for the propagation of underground EM signals \cite{Lytle1974}. {     The electrical characteristics of the earth, such as conductivity and dielectric constant have a deceitful effect on the propagation of EM signals. The values of conductivity and dielectric constant significantly influence the selection of antenna, size of the ground screen, transmission loss, phase shift, dispersion, and antenna efficiency. Therefore, conductivity and dielectric constant for various types of soils and rocks were experimentally measured in \cite{Lytle1974}.} The authors in \cite{Lytle1976} experimentally measured the conductivity of the earth surface at EM frequencies of 3 to 50 MHz at Yosemite national park. William \textit{et. al} measured high-frequency electromagnetic radiations in the borehole by neglecting the reflections and refractions from the subsurface \cite{William1982}. {    The path loss model defined in  \cite{William1982} is given as
\begin{equation}
\frac{P_r}{P_t} = \frac{f(\theta)G_t A_r \exp^{-2\alpha d}}{4 \pi r^2},
\end{equation}
where $P_r$ and $P_t$ are the received and transmitted power, respectively, $f(\theta)$ is the effective elevation radiation pattern, $G_t$ is the gain of transmitting antenna, $A_r$ is the effective area of receiving antenna, $\alpha$ is the attenuation factor, and $d$ is the distance between the transmitting and the receiving antenna.}
In \cite{harrison1990air}, the authors examined EM waves for borehole communications where a data rate of 1 bps was achieved for average conductivity and without using the repeaters while with the use of repeaters, the data rate can reach up to 100 bps. In \cite{Zheng2008}, the authors have shown the impact of soil properties, water content, network topology, and antenna type for the EM waves in the frequency range of 1 to 3 MHz for underground wireless sensor networks (UGWSNs). {    It was shown numerically in \cite{Zheng2008} that the path loss and attenuation increases with increase in the humidity and the operating frequency.} Consequently, in \cite{Silva2009}, experiments were conducted in subsoil and topsoil at 300-500 MHz frequencies for buried sensors. Furthermore, attenuation of EM signal for measurement while drilling (MWD) telemetry system was investigated in \cite{schnitger2009signal} where the maximum transmission range of 15,000 feet was achieved without using repeaters. Based on the Friis free space path loss model, the authors in \cite{VURAN2010} provided the formula for the received power in the soil medium as 
\begin{equation}
P_r \text{(dB)} = P_t \text{(dB)} + G_t \text{(dB)} + G_r \text{(dB)} - L_s \text{(dB)},
\end{equation}
where $P_t$ represents the transmit power, $G_t$ and $G_r$ are the transmit and receive antenna gains, respectively, and $L_s = L_f + L_u$ is the path loss in soil medium. $L_f$ and $L_u$ are the free space and underground  path loss, respectively. Underground path loss $L_u$ is calculated by considering the EM waves propagation characteristics in soil, such as operating wavelength and frequency, scattering, and delay distortion. Hence, $L_u = L_{\alpha} + L_{\beta}$, where $L_{\alpha}$ and  $L_{\beta}$ are {   attenuations} due to transmission loss and wavelength difference of EM signal in soil compared to air,  respectively. Therefore, $L_s$ is represented in $dB$ as follows
\begin{equation}
L_s = 6.4 +20 \log(d)+ 8.69 \alpha d+ 20 \log(\beta),
\end{equation}
where $d$ is the Euclidean distance, $\alpha$ is the attenuation constant, and $\beta$ is the phase shift constant. Both the attenuation and phase shift constants depend on the dielectric properties of the soil. The dielectric properties of the soil are calculated by using the Peplinski principle as follows \cite{Peplinski1995}:
\begin{equation}
\epsilon_s = \epsilon_r - j\epsilon_i,
\end{equation}
where $\epsilon_s$ is the complex dielectric constant of soil and water mixture consisting of  a real part $\epsilon_r$ and an imaginary part $\epsilon_i$, respectively. The real part of $\epsilon_s$ is given as
\begin{equation}\label{eq: sigmaf}
\epsilon_r = 1.15 \left(1+\frac{\rho_b (\epsilon_x^{\bar{\alpha}})}{\rho_x}+m_v^{\bar{\beta}}\epsilon_f^{\bar{\alpha}}-m_v\right)^{\frac{1}{\bar{\alpha}}}-0.68,
\end{equation}
where $\rho_b$ is the bulk density, $\rho_s = 2.66$ is density of solid soil, $\bar{\alpha} = 0.65$, $m_v$ is the water volume fraction, and $\bar{\beta} = 1.2748 - 0.519S - 0.152 C$ is the empirically determined constants for soil type. The terms $S$ and $C$ represents mass fractions of sand and clay, respectively and their values lies between 0 and 1. The effective conductivity $\epsilon_f$ in \eqref{eq: sigmaf} is given as
\begin{equation}
\epsilon_f = \frac{\epsilon_0-\epsilon_\infty}{1+(2\pi f \tau)^2}+\epsilon_\infty,
\end{equation}
where $\epsilon_0 = 80.1$ is the static dielectric constant, $\epsilon_\infty = 4.9$ is the high frequency limit, $\tau$ is the relaxation time of water, and $f$ is the operating frequency \cite{Hallikainen1985}. Similarly the imaginary part $\epsilon_i = (m_v^{\tilde{\beta}}\epsilon_f^{\bar{\alpha}})^{\frac{1}{\bar{\alpha}}}$, where $\tilde{\beta} = 1.33797 - 0.603S -0.166C$. Consequently, the attenuation constant $\alpha$ is given as
\begin{equation}
\alpha = 2 \pi f \Bigg(\frac{\mu \epsilon_r}{2}\bigg(\sqrt{1+\left(\frac{\epsilon_i}{\epsilon_r}\right)^2}-1\bigg)\Bigg),
\end{equation}
where $\mu$ is the magnetic permeability. Similarly, the phase shift constant $\beta$ is found as
\begin{equation}
\beta = 2 \pi f \Bigg(\frac{\mu \epsilon_r}{2}\bigg(\sqrt{1+\left(\frac{\epsilon_i}{\epsilon_r}\right)^2}+1\bigg)\Bigg).
\end{equation}
It is clear from the expression of both the attenuation and phase shift constants that the propagation loss of EM depends on the operating frequency, soil composition, water content, and bulk density. Furthermore, the authors in \cite{VURAN2010} have investigated the path loss in the presence of two paths between the transmitter and the receiver. The authors have neglected the second path effect in high depth scenarios due to no reflection from the ground surface while for low depth scenario two-path model was considered which is given in $dB$ as follows:
\begin{equation}
L_t = L_s - L_v,
\end{equation}
{    where $L_v$ correspond to the second path loss, given as}
\begin{eqnarray}
L_v^2 &=& 1 + \left(\gamma \exp^{{(-\alpha\Delta(r))}^2}\right)-2\gamma \exp^{(-\alpha \Delta(r))} \nonumber \\
& & \cos \bigg(\pi - \left(\phi-\frac{2\pi \Delta(r)}{\lambda}\right)\bigg),
\end{eqnarray}
where $\gamma$ and $\phi$ are the amplitude and phase reflection coefficients, respectively, $\lambda$ is the wavelength, and $\Delta(r) = r -d$ is the difference between the two paths. Based on the above channel model, the authors in \cite{Silva2010b} proposed a testbed for UGWSNs. The authors in \cite{Yoon2011} also compared the theoretical and measured results for UGWSNs where the above analytical model fits well within 3.45 dBm of the measured data. 
In \cite{ Akkas2012} EM waves in Terahertz band (0.1-120 THz) were investigated for oil reservoirs. {    The path loss (in dB) for EM signals in the Terahertz band is given in \cite{ Akkas2012} as follows
\begin{equation}
L_{tot} = L_{sp} + L_w + L_o,
\end{equation} 
where $L_{sp}$ is the spreading loss, $L_w = k_w(f)d$ is the absorption loss due to water, and $L_o= k_o(f)d$ is the absorption loss due to oil. The spreading loss $L_{sp}=( \frac{4 \pi f d}{c})^2$, where $d$ is the distance, $f$ is the operating frequency,  $c$ is the speed of light, and $k_w(f),~k_o(f)$ are the absorption coefficients of water and oil, respectively.}
Although the THz band provides high capacity for UGWSNs, their range is limited to few centimeters. Hence, the concept of low frequency (below 500 {   kHz})  was introduced in \cite{Yoon2012, Yoon2012a, Ghazanfari2012} to achieve more considerable transmission distance (in tens of meters) for UGWSNs. {     The received power formula for the EM signals in \cite{Yoon2012, Yoon2012a, Ghazanfari2012} is given as
\begin{equation}
P_r = \kappa \frac{\exp^{-2 \alpha d}}{d^2},
\end{equation}
where $\kappa = \frac{A_r \cos \theta}{2 \eta}(\frac{I \mu_0 \omega}{4 \pi})^2$, $\theta$ is the phase angle, $\eta$ is the intrinsic wave impedance, $\mu_0$ is the permeability of air, $\omega$ is the angular frequency, and $I$ is the current.
}

The impact of the carrier frequency, transmission distance, depth, and modulation type was experimentally tested in \cite{Goyal2014, Yu2015, Horvat2016} for UGWSNs. The optimum frequency range in 10-100 MHz was identified in \cite{Jiang2012} for energy harvesting in UGWSNs. To reduce the battery consumption and to improve the signal to noise ratio (SNR) at the receiver, a code division multiplexing scheme was proposed in \cite{Hikaru2016}.
Consequently, the impact of soil type on multi-carrier modulation was examined in \cite{Salam2016} which showed that the data rate of 124 Mbps is achievable for the transmission distance of 12 m for IoUT. In \cite{Salam2017}, the authors used pulse amplitude modulation, quadrature phase shift keying, m-ary quadrature amplitude modulation, and Gaussian minimum shift keying for IoUT.
Moreover, it was shown in \cite{Salam2017} that adaptive equalization improve the performance of the underground channel. Furthermore, the authors in \cite{Suherman2018} tested 97-130 MHz EM frequencies for underground radio propagation by using the same model in \cite{VURAN2010}; however, the error was almost 50 \% at such high frequencies. In \cite{SALAM2019}, a real-time soil moisture sensing and permittivity estimation system called Di-Sense was proposed to implement IoUT for agricultural applications. Software defined radio-based experiments were conducted where the permittivity and soil moisture were calculated at a depth of 4 cm and horizontal distance of 1 to 15 m, for the frequency range of 100-500 MHz. 

Recently, the authors in \cite{Zemmour2017} investigated the soil effects, the orientation of the buried antenna, and depth on the underground to the above ground wireless communication link. {     The empirical path loss for underground to above ground link is given in logrithmic scale  as follows \cite{Zemmour2017}
\begin{equation}
L_{ag} = -147.6 + 20 \log d_{ag} + 20 \log f
\end{equation} 
where $d_{ag}$ is the path length.}
Consequently,  empirical studies were conducted in \cite{ Du2017} by using the above model to show the propagation characteristics of the underground to above ground communication link at 2.4 GHz and 433 MHz, respectively. {    The emperical model in \cite{VURAN2010} and its variants are mostly employed  in the literature for EM-based underground communication. However, these emperical models ignore the physical characteristics of subsurface EM field analysis. Therefore,  recently in \cite{Salam2019a}, the author proposed an analytical solution for EM-based underground communications which rely on the Maxwell-Poynting theory. The derived expressions in \cite{Salam2019a} relate the soil properties to the electromagnetic field of the buried dipole antenna. Table \ref{Tableechnl} summarizes the literature on channel modeling for EM-based IoUT.}
\begin{table*}
\footnotesize
\centering
\caption{{    Summary of channel modeling for EM-based IoUT.}}
\label{Tableechnl}
\begin{tabular}{|p{1.8cm}|p{2.8cm}|p{2.4cm}|p{3.0cm}|p{2.4cm}|p{1.9cm}|}
\hline
\hline
\textbf{Ref.}              & \textbf{Data rate}         & \textbf{Frequency range} & \textbf{Issue addressed}& \textbf{Applications} & \textbf{Year}\\ \hline 
\cite{Wait1971} & - & 1-10 MHz & Propagation characteristics & Seismic/Agriculture & 1971\\
\cite{Sivaprasad1973} & - & - & Structure of soil effect on EM waves propagation & Seismic/Agriculture & 1973\\
\cite{Lytle1974} and \cite{Lytle1976} & - & 3-50 MHz & Electrical characteristics of soil & Seismic/Agriculture & 1974 and 1976\\
\cite{harrison1990air} & 1-100 bps & -  & EM waves for borehole communications & Oil and Gas & 1990\\
\cite{Zheng2008} & - & 1-3 MHz  & Impact of soil and network parameters & Agriculture & 1990\\
{    \cite{schnitger2009signal}}& {    -} & {    2-6 Hz}  & {    Investigation of depth on the signal strength} &{    Oil and Gas}& {    2009}\\
\cite{Silva2009} & - & 300-500 MHz  & Impact of soil type & Agriculture & 2009\\
\cite{VURAN2010} & - & -   & development of the path loss model & Agriculture & 2010\\
\cite{Silva2010b} & - & -   & Test-bed & Agriculture & 2010\\
{    \cite{Yoon2011}}& {    -} & {    0.3-1.3 GHz}  & {    Comparison of theoretical and experimental results} &{    Agriculture}& {    2011}\\
\cite{Akkas2012} & - & 0.1-120 THz  & Channel model & Oil and Gas & 2012\\
\cite{Yoon2012, Yoon2012a, Ghazanfari2012} & - & below 500 {   kHz}  & Propagation characteristics & Agriculture & 2012\\
\cite{Jiang2012} & - & 10-100 MHz  & Energy harvesting & Seismic/Agriculture & 2012\\
\cite{Goyal2014, Yu2015, Horvat2016} & - & 433 MHz  & Propagation characteristics & Agriculture & 2014-2016\\
{    \cite{Hikaru2016}}& {    -} & {    -}  & {    New transmitter and receiver configurations to improve the sensor node lifetime} &{    Seismic}& {    2016}\\
{     \cite{Zemmour2017}}& {    -} & {    3.1-10.6 GHz}  & {    Impact of soil on ultrawideband underground to above ground communication link } &{    Agriculture}& {    2017}\\
\cite{Salam2016} & 124 Mbps & 433 MHz  & Multi-carrier modulation for EM-based IoUT & Agriculture & 2017\\
{    \cite{Salam2017}}& {    -} & {    100-300 MHz}  & {    Using of the direct, reflected, and lateral components of the underground channel to improve the BER} &{    Agriculture}& {    2017}\\
{     \cite{Du2017}}& {    -} & {    433 MHz and 2.4 GHz}  & {    Influence of depth on the propagation of EM signal} &{    Agriculture}& {    2017}\\
\cite{Suherman2018} & - & 97-130 MHz  & Soil moisture sensing & Agriculture & 2018\\
{     \cite{SALAM2019}}& {    -} & {    433 MHz}  & {    Estimation of relative permittivity and soil moisture} &{    Agriculture}& {    2019}\\
{     \cite{Salam2019a}}& {    -} & {    -}  & {    Underground channel modeling by using  Maxwell-Poynting theory } &{    Agriculture}& {    2019}\\
\hline 
\hline
\hline        
\end{tabular}
\end{table*}
\begin{figure}[t]
  \centering
  \includegraphics[width=0.5\textwidth,height=0.28\textheight]{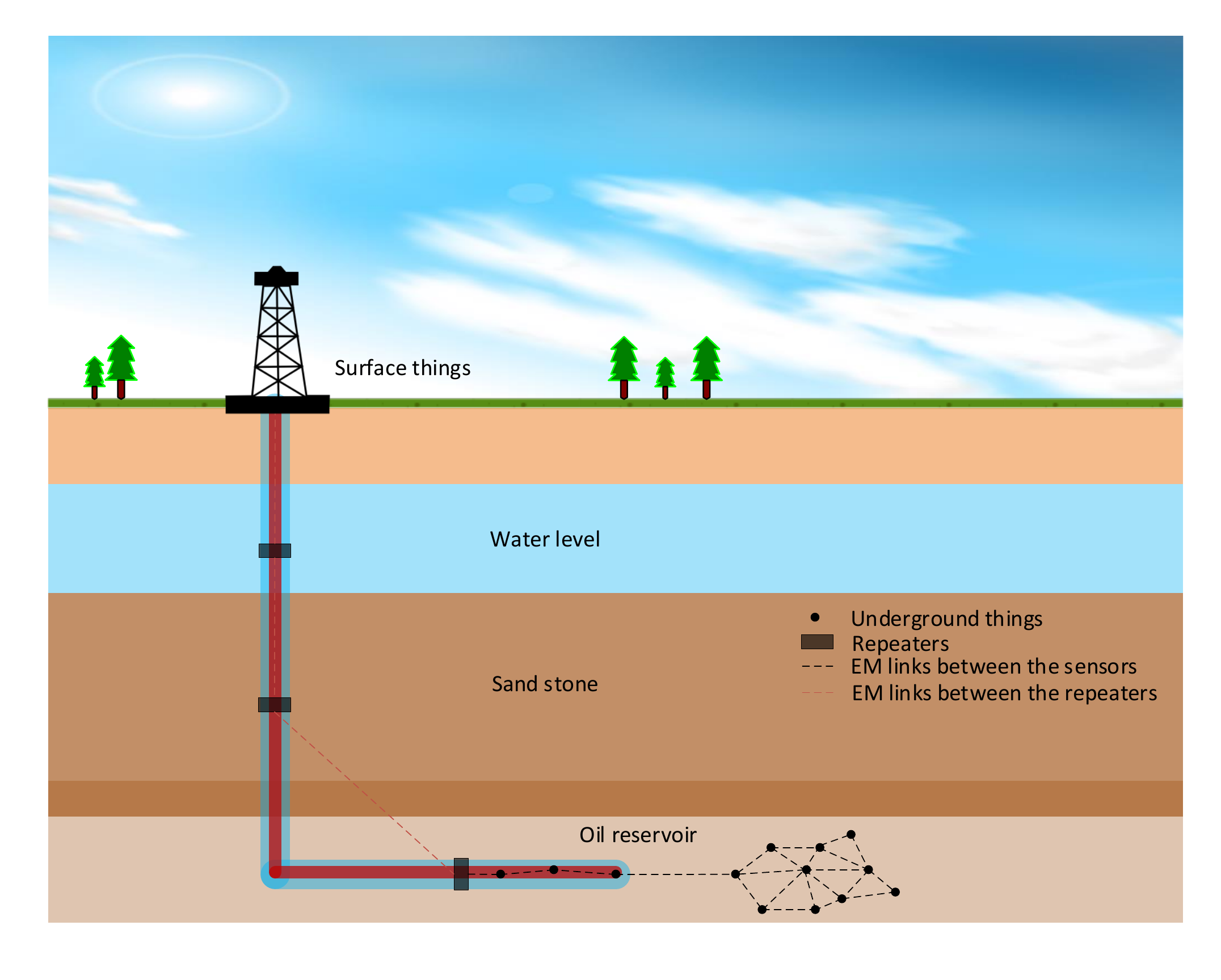}\\
  \caption{Network model for EM-based IoUT for oil and gas reservoirs.}
  \label{EMnet}
\end{figure} 

\begin{figure}[t]
  \centering
  \includegraphics[width=1\columnwidth,height=0.25\textheight]{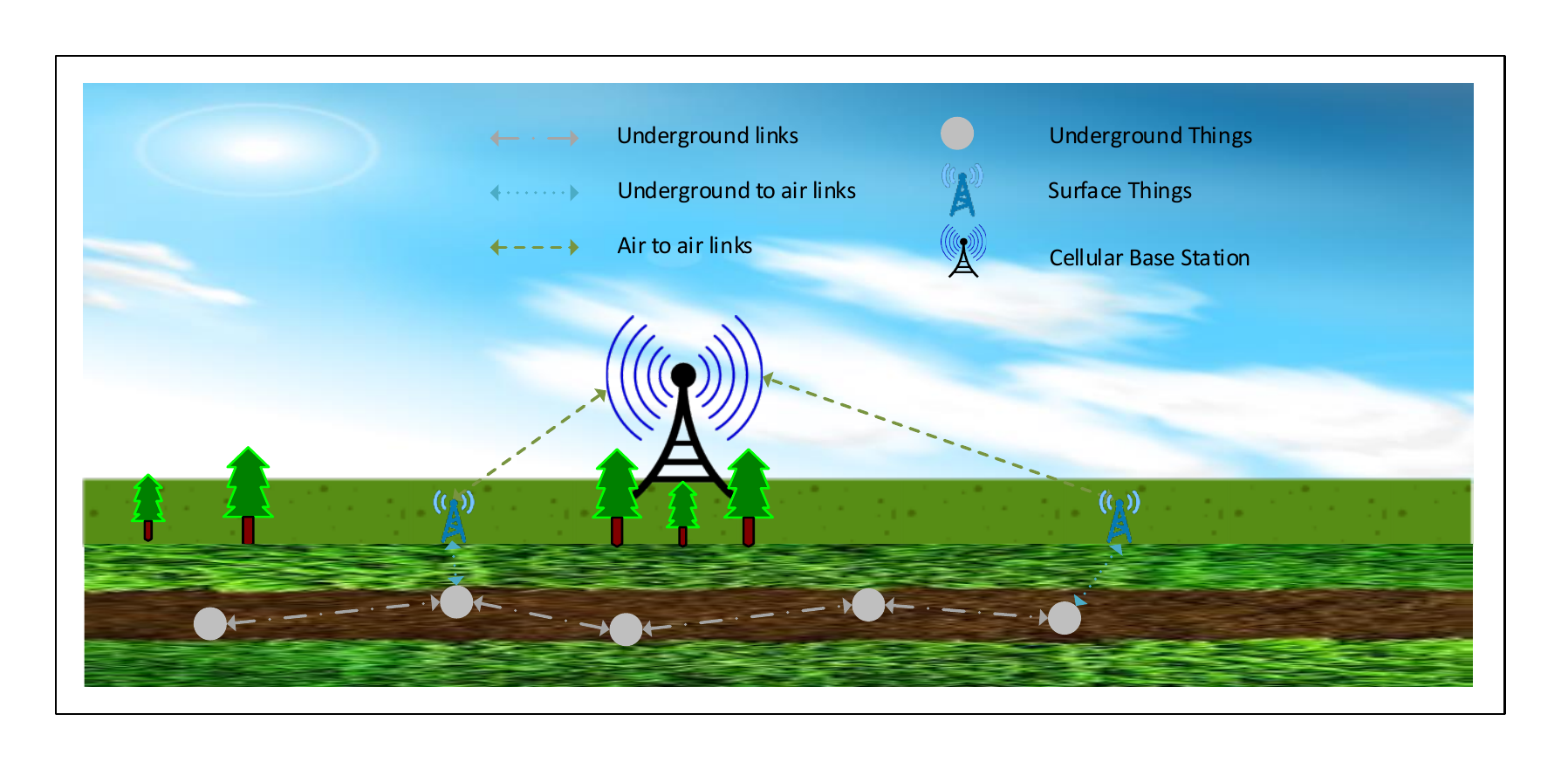}\\
  \caption{Network model for EM-based IoUT for agricultural applications.}
  \label{EMnet2}
\end{figure}

\subsection{Networking}
The literature on channel model for EM-based IoUT is rich. However, few works exist on the routing protocols. In this section, we cover the existing research on networking layer protocols for  EM-based IoUT.
{    Fig. \ref{EMnet} and \ref{EMnet2}  show} examples of a multi-hop network for an EM-based oil and gas IoUT and agricultural IoUT, respectively. Due to the limited transmission range of EM waves in the harsh underground environment multi-hop communication is well-suited. For example, the idea of dense sensor networks with multi-hop communication for oil and gas exploration was presented in \cite{Savazzi2013}. Moreover, the authors in \cite{Conceicao2014}  developed a TCP/IP based simulator for the IoUT.
Furthermore, they evaluated the performance of various multiple access schemes in \cite{Conceicao2016}. The path connectivity problem for EM-based IoUT was investigated in \cite{Dung2016} which showed that low volumetric water content and low operating frequency lead to a higher probability of connectivity. In \cite{Liu2016}, the throughput of EM-based IoUT was optimized to achieve the QoS requirement. Recently, the influence of soil texture, particle density, and bulk density on the hop count was examined for IoUT where the number of hops between a source and a sink increases with an increase in the water content and clay in the soil. A relay based approach with physical constraints on the relay location, propagation environment, and load balancing was examined in \cite{YUAN2017} to improve the lifetime of IoUT.

\section{Acoustic Waves for IoUT}
Most of the communication and detection techniques for underground measurements are based on acoustic waves.   Geologists use acoustic waves to look for underground resources, such as oil and gas. Acoustic waves are transmitted into the ground, and the reflection is measured from the propagation of the acoustic waves. Moreover, acoustic waves are used in drilling to communicate with underground equipment. {     Major applications of acoustic waves-based IoUTs include smart seismic exploration, earthquake monitoring, buried pipeline monitoring, and smart drilling for oil and gas reservoirs.} 
The research work on acoustic-based underground communications systems is rich and can support many other applications. Based on the signal generation, the acoustic-based methods can be broadly classified into passive and active type methods. 
In passive acoustic-based methods, the subsurface environment generates an acoustic signal  caused by natural events, such as earthquakes, nuclear explosion, and volcanic explosions. In such circumstances, the sensors are placed in the vicinity of the event area. These sensors detect the infrasonic signals which help in the prediction of a natural disaster. Moreover, sudden changes underground, such as rock crack formation, structural transformation, and pipeline leakage can also be detected by using passive acoustic methods.
\begin{figure}[t]
  \centering
  \includegraphics[width=0.5\textwidth,height=0.25\textheight]{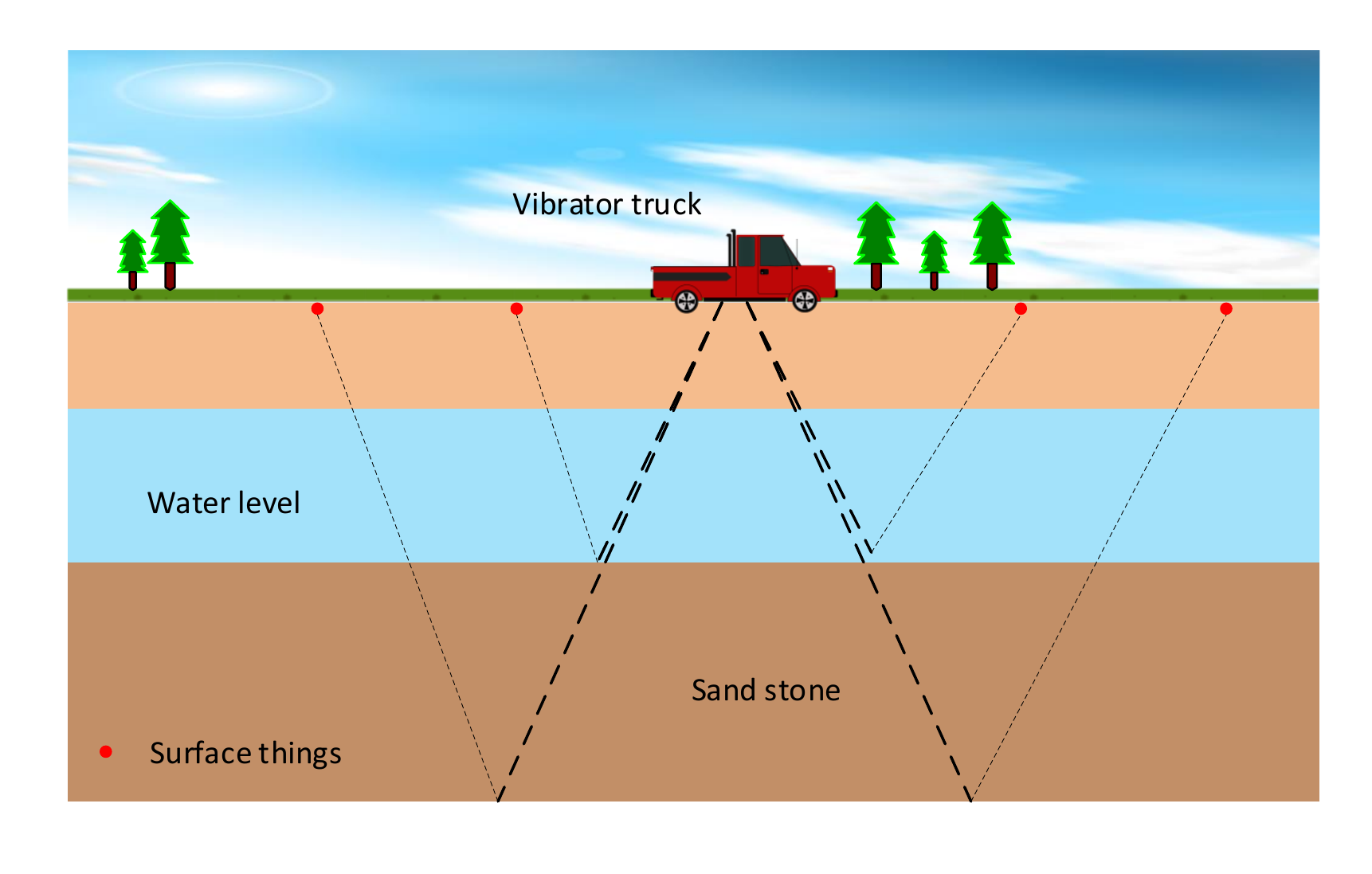}\\
  \caption{Architecture of an active acoustic-based IoUT for seismic monitoring.}
  \label{acnet2}
\end{figure}
In active acoustic-based methods, the signal is generated by an artificial explosion or vibration which is sent underground to estimate the properties of the earth’s subsurface (see Fig.~\ref{acnet2}). The popular application of such method is reflection-based seismology.
Due to the low propagation speed of acoustic waves, they are mostly used for detection purposes in soil rather than for communication.
In \cite{Oelze2002}, the authors investigated the speed of sound in the soil. Acoustic signals were transmitted through different samples of soil and received by the hydrophones. The attenuation coefficients were calculated for the frequencies range {    from} 2 to 6 {   kHz}. The proposed empirical solution was able to detect an object buried at 40 cm. Similarly, in \cite{Adamo2003} and \cite{Adamo2004}, soil moisture was measured by using the speed-moisture curves for underground acoustic signals transmission. Moreover, acoustic waves with a frequency of 900 {   Hz} was used in \cite{sharma2010continuous} to estimate the moisture content of the soil. A universal soil loss equation was derived in \cite{freire2015sound} for acoustic waves propagation in soil at 16 {   kHz} frequency. {     The soil loss factor due to the erosion is expressed  in \cite{freire2015sound} as $L_{e_s} = \rho e_s \tau c_m \varrho$, where $\rho$ is the soil loss due to rain factor, $e_s$ is the soil erosion factor, $\tau$ is the topographic factor of soil, $c_m$ is the cover management, and $\varrho$ is the support practices factor.
} Recently, the authors in \cite{Singer2017} and \cite{Sijung2018} proposed an acoustic based wireless data transmission system (SoilComm) for IoUTs. SoilComm system was able to transmit the sensing data over 30 m distance through soil.

%
%
\begin{table*}
\footnotesize
\centering
\caption{{    Comparison  of acoustic waves-based IoUT.}}
\label{Tableacousticdown}
\begin{tabular}{|p{1.3cm}|p{1.2cm}|p{2.4cm}|p{3.5cm}|p{2.5cm}|p{1.4cm}|}
\hline
\hline
\textbf{Ref.}           & \textbf{Data rate} & \textbf{Depth} &  \textbf{Issue addressed}& \textbf{Applications} & \textbf{Year}\\ \hline 
{    \cite{Oelze2002}} &  -  & - & {    Soil sampling} & {    Agriculture}& {    2002}\\
{    \cite{Adamo2003, Adamo2004}} &  -  & - & {    Soil moisture detection} & {    Agriculture}& {    2003-2004}\\
\cite{gardner2006acoustic} &  20 bps &1120 m& Down-hole communication & Underground drilling & 2006\\
\cite{neff2007field} &  20-60 bps &1000 m& Field tests for down-hole communication & Underground drilling & 2007\\
{    \cite{sharma2010continuous}} &  -  & - & {    Soil moisture detection} & {    Agriculture}& {    2010}\\
{    \cite{KHAN2010}} &  -  & - &  {    Detection of mines using acoustic waves} & {    Underground mines detection}& {    2010}\\
{    \cite{Kang2011}} &  -  & - &  {     Detection of rock deformation by using acoustic emission } & {    Seismic} & {    2011}\\
\cite{Gutierrez2013} &  6 and 20 kbps & 55 and 4.5  m& OFDM for down-hole communication & Underground drilling & 2013\\
\cite{Wei2013} &  - & -  & Impact of pipe joints on signal transmission  & Underground drilling & 2013\\
\cite{Ahmad2014} &  - & -& Impact of multi-phase flow with ASK and FSK  & Underground drilling & 2014\\
\cite{Pelekanakis2014} &  400 bps & 1000 m & Trellis coded modulation for down-hole communication & Underground drilling & 2014\\
{    \cite{freire2015sound}} &  -  & - &  {    Universal soil loss equation} & {    Agriculture}& {    2015}\\
{    \cite{Sun2010A,VanHieu2011,Su2015}} &  -  & - &  {    Detecting cracks in pipelines} & {    Underground pipelines monitoring}& {    2011-2015}\\
{    \cite{Alenezi2017}} &  -  & - &  {    Investigation of single channel and multi-channel accelerometers } & {    Down-hole telemetry} & {    2017}\\
\cite{MA2018} &  500 bps & 53.76 m & NC-OFDM for down-hole communication & Underground drilling & 2018\\
{    \cite{Singer2017, Sijung2018}} &  -  & - &  {    Wireless data transmission in soil} & {    Agriculture} & {    2018}\\
\hline 
\hline
\hline        
\end{tabular}
\end{table*} 
In addition to the investigation of soil properties, acoustic waves are widely used for down-hole telemetry purposes. In acoustic telemetry, the steel walls of the drill-string are used as a source of a communication channel. Acoustic-based telemetry system consists of a piezoelectric-electric transmitter underground, repeater at 500-2000 m apart, and a transceiver at the ground surface.  {     The down-hole transmitter encode the sensor's data and convert it into the acoustic signal which then propagates to the surface by the drill string. The drill string consists of pass-bands and stop-bands in the channel. It was found in \cite{gardner2006acoustic} that attenuation on the drill-string is in the range of 4 to 7 dB per 1000 feet. The expression for the capacity of the acoustic telemetry system was calculated in \cite{gardner2006acoustic} as follows
\begin{equation}
C_a = \int \log_2 \left(1+\frac{P_{s}(f)}{P_{N_d}(f)+|H(f)|^{-2}P_{N_s}(f)}\right) df
\end{equation}
where $P_s$, $P_{N_s}(f)$ and $P_{N_d}(f)$ are the power spectra of the signals, surface noise and drilling noise, respectively. The transfer function for the acoustic channel is represented by $H(f)$.
} Acoustic waves passing through the drill string are highly attenuated and therefore require a sufficient number of repeaters. Additionally, the drilling noise also affects the propagation of acoustic waves along the string. The authors in \cite{gardner2006acoustic} were able to achieve the data rate of 20 bps at a depth of 3695 feet. In \cite{neff2007field}, field tests were performed by using acoustic telemetry where data rates of 20, 40, and 60 bps were achieved at a depth of 1000 m. In \cite{farraj2012channel} and \cite{farraj2012acoustical}, a testbed was developed to study the channel behavior for acoustic waves over the string pipes. The results in \cite{farraj2012channel} and \cite{farraj2012acoustical} have shown that acoustic waves suffer from noticeable dispersion and pipe strings act as a frequency selective channel.

Authors in \cite{Gutierrez2013} also performed experiments by using acoustic waves for downhole communications where the data rate of 20 and 6 kbps were achieved for 4.5 and 55 m depth, respectively. Acoustic waves were generated by using a magnetostrictive actuator which converts electrical signals into acoustic vibrations. The acoustic signals were then transmitted over the drill string to the bottom and received back at the surface by the geophones. For a frequency selective channel of the drill string, orthogonal frequency division multiplexing (OFDM) was used. {    The drill string was modeled as a series of alternating short and long resonators where each resonator is described by a scattering matrix $\mathbf{S}$. The string is considered as a two port device where $\mathbf{S}_{11}$ and $\mathbf{S}_{22}$ represents the reflections while $\mathbf{S}_{12}$ and $\mathbf{S}_{21}$ measure the transmission of the acoustic signal, respectively. The scattering parameters are obtained as \cite{Gutierrez2013}
\begin{equation}
\mathbf{S}_{11} = \mathbf{S}_{22} = \dot{r} \left(1- \frac{(1-\dot{r}^2)\exp^{-2j\gamma \ell}}{1-\dot{r}^2\exp^{-2j\gamma \dot{r}}}\right),
\end{equation}
and
\begin{equation}
\mathbf{S}_{12} = \mathbf{S}_{21} = 1- \frac{(1-\dot{r}^2)\exp^{-2j\gamma \ell}}{1-\dot{r}^2\exp^{-2j\gamma \ell}},
\end{equation}
respectively. The terms $\ell$ and $\dot{r}$ represents the length of the segment and the reflection coefficient, respectively, while $\gamma$ is the function of the attenuation coefficient $\alpha$, frequency $f$, and velocity of the acoustic signal $v$ as,  $\gamma = \frac{2 \pi f}{v}-j \alpha$. Based on the sub-matrices of $\mathbf{S}$, the $\mathbf{T}$ matrix is written as 
\begin{equation}
\mathbf{T} = \frac{1}{\mathbf{S}_{12}}\begin{bmatrix}
\mathbf{S}_{12}\mathbf{S}_{21} & \mathbf{S}_{11}\\
-\mathbf{S}_{22}&1
\end{bmatrix}.
\end{equation}
The channel frequency response for the whole string is then calculated as
\begin{equation}
\mathbf{T}_s = \prod_{i=1}^N \mathbf{T}_i
\end{equation}
where $i=1,2,...N$ {   is} the number of segments of the drill string.
}
Besides the modeling of drill pipe,  arrangements of pipes also play an important role in acoustic communication during drilling. Hence, the authors in \cite{Kumar2013} argued that ascend-to-descend arrangement of pipes provide best telemetry performance for downhole acoustic communication. The problems of acoustic noise and attenuation of the acoustic signal due to the pipes joints {   were} studied in \cite{Wei2013}. The authors proposed a single carrier with frequency domain equalization (SC-FDE) to improve the reliability of the acoustic transmission along the pipe strings. The impact of multiphase flow was examined in \cite{Ahmad2014} for downhole acoustic communication with amplitude shift keying and frequency shift keying modulation schemes. Moreover, the authors in \cite{Pelekanakis2014} introduced the use of trellis coded modulation for downhole acoustic communication where a more realistic model of 1000 m depth was considered with an achievable data rate of up to 400 bps.
\begin{figure}[t]
  \centering
  \includegraphics[width=0.5\textwidth,height=0.15\textheight]{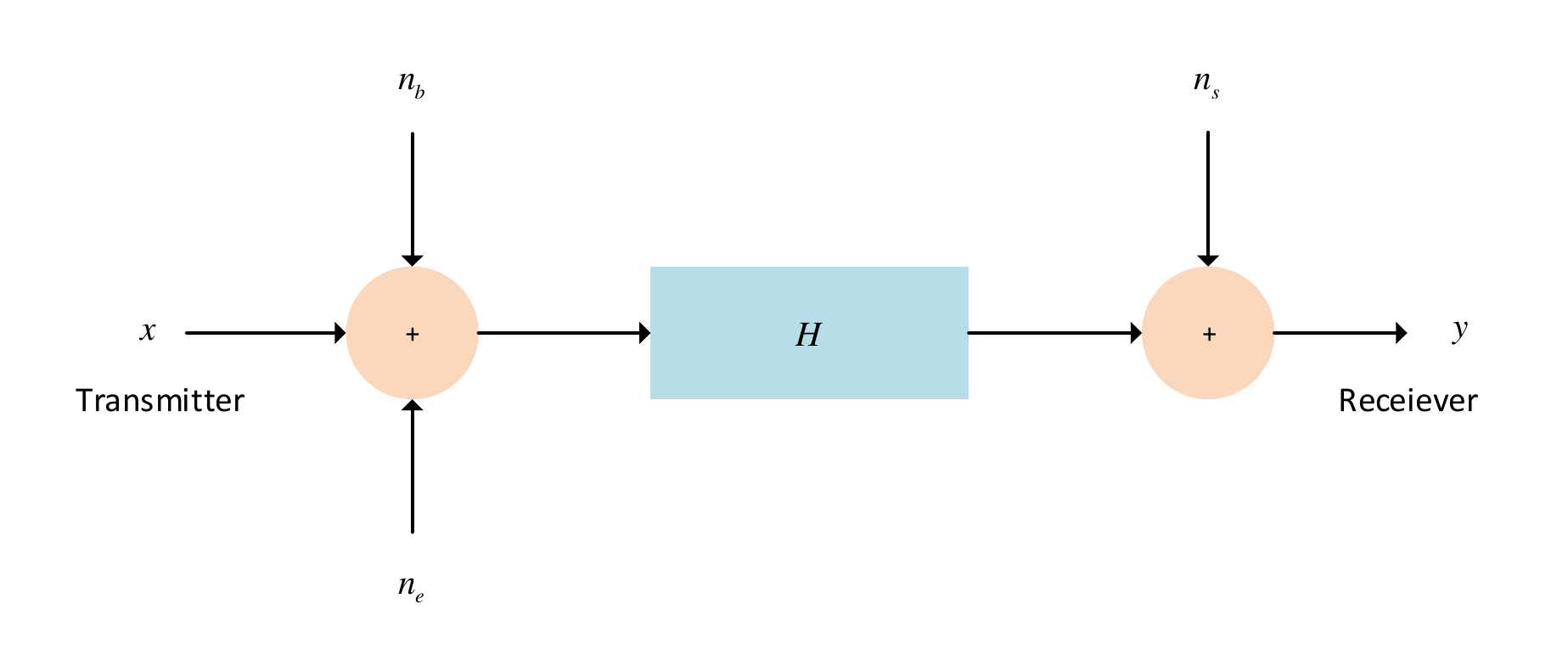}\\
  \caption{Equivalent transmission model of the drill string.}
  \label{drill_string}
\end{figure}

{    
Recently, non-contiguous OFDM with adaptive pilot design was used in \cite{MA2018} to provide data rate of up to 500 bps at the depth of 53.76 m. The drill string was modeled to the schematic diagram shown in Fig.~\ref{drill_string} where $x$ is the transmitted signal, $y$ is the received signal, $H$ is the channel transfer function, $n_b$ is drill bit noise, $n_e$ is the environment noise, and $n_s$ is the surface noise. Then the received signal is written as
\begin{equation}
y = Hx+H(n_b + n_e)+n_s = H(x+n_d)+n_s,
\end{equation} 
where $n_d$ is the total transmitter noise. Least square estimation was used in \cite{MA2018} to estimate the acoustic channel characteristics along the drill string.
}

Moreover, in addition to the channel model for acoustic waves based underground communications, work on the transceivers design for such applications has also been an active area of research. For example, in \cite{Li2016}, a novel receiving unit for acoustic communication along the drill string was proposed. Similarly, a tri-axial accelerometer was used in \cite{Alenezi2017} to compare the single channel and multi-channel uphole acoustic communication in oil wells. Nevertheless, Gao \textit{et. al}  studied the transmission of acoustic waves along the drill strings for various applications \cite{gao2018study}.
Besides, the literature on transmission characteristics and transceiver design, studies on the characteristics of acoustic signals also exists for rock failure \cite{Kang2011}, cracks in pipelines \cite{Sun2010A,VanHieu2011,Su2015}, and landmines detection \cite{KHAN2010}.  Table {   \ref{Tableacousticdown}} summarizes the literature on acoustic-based IoUT.

\section{Mud Pulse Telemetry for IoUT}
\begin{figure}[t]
  \centering
  \includegraphics[width=0.5\textwidth,height=0.28\textheight]{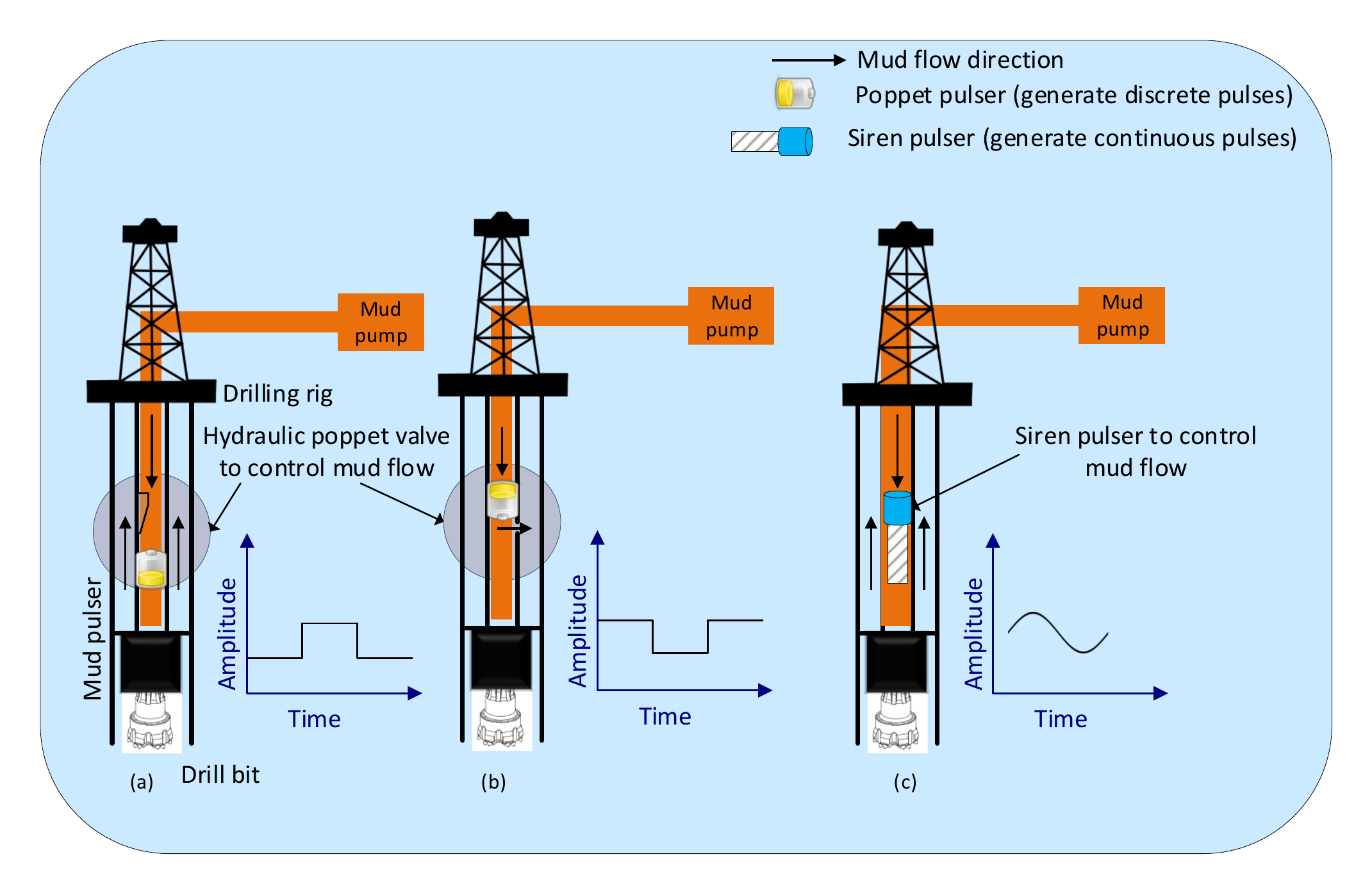}\\
  \caption{Three different types of MPT systems. (a) Positive pulses from blocking/unblocking of the fluid; (b) Negative pulses by pressure in the drill string; (c) Continuous pulses by using a rotor.}
  \label{mpt}
\end{figure}

The most common and mature method for downhole communication is mud pulse telemetry (MPT) which dates back to the mid of the $\text{19}^{\text{th}}$ century. The early MPT systems were able to only communicate the azimuth and inclination information for the wells navigation. The main concept of MPT lies in the circulation of the mud for the transmission of the data \cite{thakur2018most}. During the drilling process, the pumps at the surface circulate the mud down to the drill string through the pressure pulses \cite{hutin2001new, Klotz2008}. The mud is used to cool the downhole drill string components, carry information from the bottom to the surface, and balance the pressure. The mud passes through a valve which restricts and generates the pressure waves. The pressure pulses are controlled and are used to modulate, frequency, amplitude, and phase of the mud pulse signals \cite{hahn2008reciprocating}. Three different types of mud pulse signals are transmitted; i.e., positive, negative, and continuous wave pulses  as shown in Fig. \ref{mpt} \cite{QU2018}. The signal processing modules at the surface recognize these pressure pulses. The pressure pulse signals in MPT systems are encoded by various techniques to carry the critical information, such as temperature, pressure, and conductivity etc. of the well. {   Table \ref{Tablemudpulse} summarizes the literature on MPT systems based IoUT.} Although the MPT systems are mature, the mud pulse signals suffers from several impairments which are discussed in the following subsections.
\subsection{Mud Pump Noise}

During the down-link transmission of the mud pulse signal, the piston in the valve moves {    back  and forth} to generate the signal. At the same time the up-link signal is generated in the similar fashion in opposite direction resulting in the interference between the down-link and up-link signals \cite{Jianhui2007}. Moreover, the pressure of the pump creates noticeable amount of frequency and amplitude varaitions in the frequency range of 1-20 Hz. To diminish these effects, MPT systems uses two different transducers at the surface receivers which are well spaced \cite{wasserman2008mud, Wassermann2009}. In \cite{reckmann2014downhole} and \cite{jarrot2018wireless}, the authors have used least mean square filtering algorithm to reduce the noise generated by the mud pumps.

\subsection{Attenuation and Dispersion}

As the mud pulse signals propagate along the borehole, the signals are attenuated and dispersed due to under-balanced drilling mud \cite{harrell2000method}. The major sources of attenuation are the mud type, joints in the drill string, signal frequency, diameter of the string, and borehole depth \cite{lin2013calculation}. Low frequency signals can be used to avoid excessive attenuation of the mud pulse signals \cite{hutin2016zero}.

\subsection{Rock Fragments and Gas Leakage}
During the drilling process, rock particles and gas may enter the mud used for the pressure pulses which changes the density and compressibility of the mud \cite{lu2013hard}. These rock particles and gas reduce the transmission speed of the pressure pulse waves. Hence, it is essential to study the velocity continuity of the drilling mud because the gas leakage into the mud can lead to unstable drilling which can cause environmental pollution and potential loss of human lives \cite{Mwachaka2018}.

\begin{table*}
\footnotesize
\centering
\caption{{    Summary of the literature  on MPT systems.}}
\label{Tablemudpulse}
\begin{tabular}{|p{1.3cm}|p{1.2cm}|p{2.4cm}|p{3.5cm}|p{2.5cm}|p{1.4cm}|}
\hline
\hline
\textbf{Ref.}           & \textbf{Frequency} & \textbf{Depth} &  \textbf{Issue addressed}& \textbf{Applications} & \textbf{Year}\\ \hline 
{    \cite{hutin2001new}} &  {    10-12 Hz}  & {     1.7 {   km}} & {     Generation, transmission, and reception of mud pulse signals} & {    Deep water drilling}& {    2001}\\

{    \cite{Klotz2008}} &  {    30 Hz}  & {     0.5 {   km}} & {     Novel mud pulser which handle the varying nature of the channel} & {    Oil and Gas reservoirs}& {    2008}\\

{    \cite{hahn2008reciprocating}} &  {    40 Hz}  & {    -} & {     Novel method by using a linear actuator to generate pressure pulses} & {    Down-hole telemetry}& {    2008}\\

{    \cite{QU2018}} &  {    12-24 Hz}  & {    150 m} & {     Adaptive noise cancellation technique for the mud pump} & {    Underground drilling}& {    2018}\\

{    \cite{Jianhui2007}} &  {     1-20 Hz}  & {    -} & {     Novel decoding technique to overcome the pump noise, reflection noise, and random noise for MPT systems } & {    Underground drilling}& {    2007}\\

{    \cite{reckmann2014downhole}} &  {    -}  & {     -} & {     Down-hole noise cancellation} & {    Underground drilling}& {    2008}\\

{    \cite{harrell2000method}} &  {    -}  & {     -} & {     Novel MPT system for under-balanced drilling} & {    Underground drilling}& {    2000}\\

{    \cite{lin2013calculation}} &  {     10-100 Hz}  & {     -} & {     Investigation of the pressure wave propagation characteristics} & {    Oil and Gas exploration}& {    2013}\\

{    \cite{hutin2016zero}} &  {     -}  & {     -} & {     Method to detect increase or decrease in the pressure for the MPT systems} & {    Oil and Gas exploration}& {    2016}\\

{    \cite{lu2013hard}} &  {     10-1000 Hz}  & {     -} & {     Novel hard rock drilling technique by using abrasive water} & {    Underground drilling}& {    2016}\\

\hline 
\hline
\hline        
\end{tabular}
\end{table*}

\section{Magnetic Induction for IoUT}

One of the major factors which limit the evolution of IoUT is the challenging underground environment. We have previously discussed that the {   heterogeneous} soil medium and water content of the soil limit the transmission range of EM-based IoUT \cite{Silva2009}. Hence, magnetic induction (MI) has been introduced to overcome the limitations of EM waves for IoUT \cite{Sun2009, Caffrey2013}. In this section, we will cover multiple aspects of MI-based IoUT which include channel modeling, networking, and localization.
\begin{figure}
\centering
\includegraphics[width=1\columnwidth]{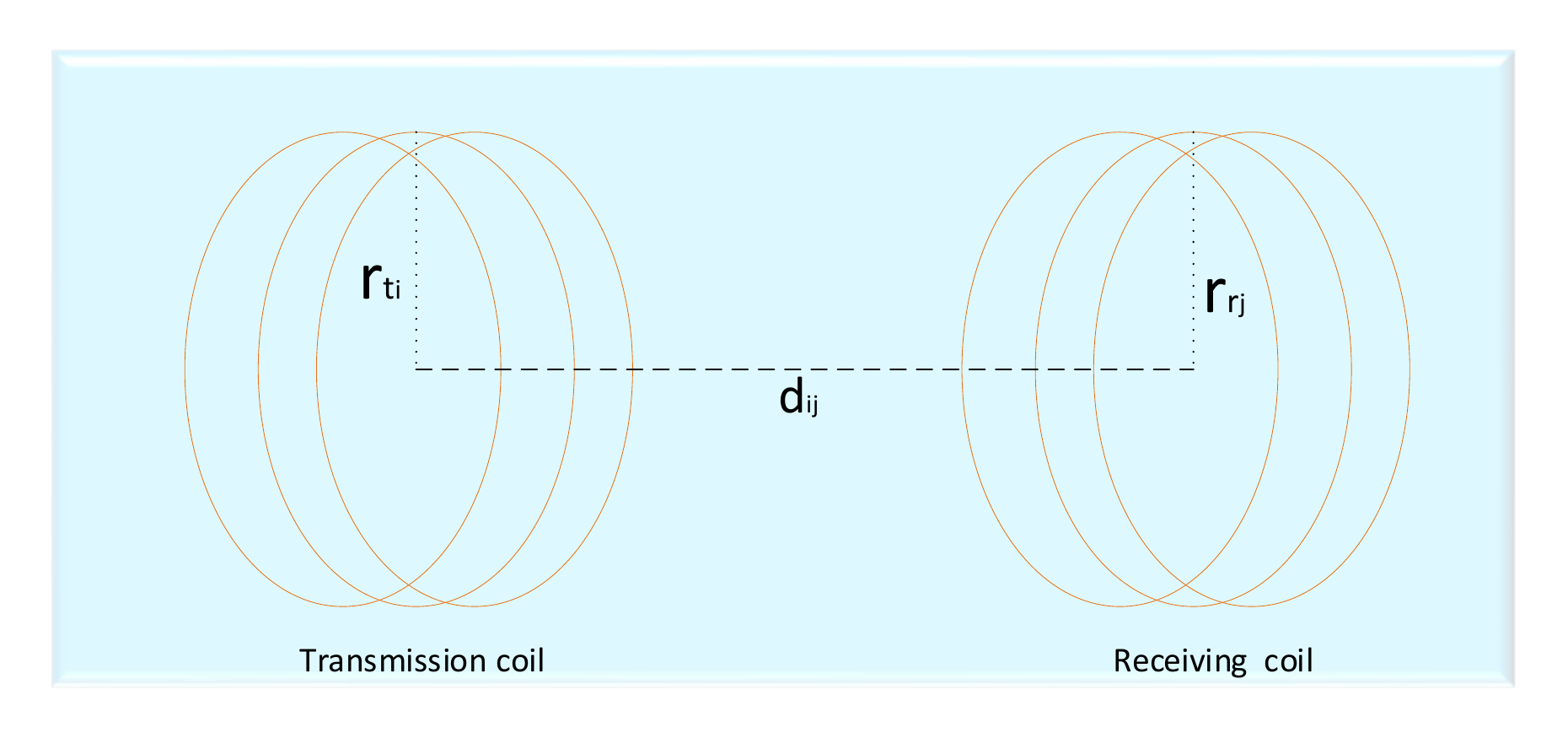}  
\caption{MI communication link.\label{fig:micoils}} 
\vspace{-1.5 em}
\end{figure}
\subsection{{    Channel Modeling}}
\begin{figure}
\centering
\includegraphics[width=1\columnwidth]{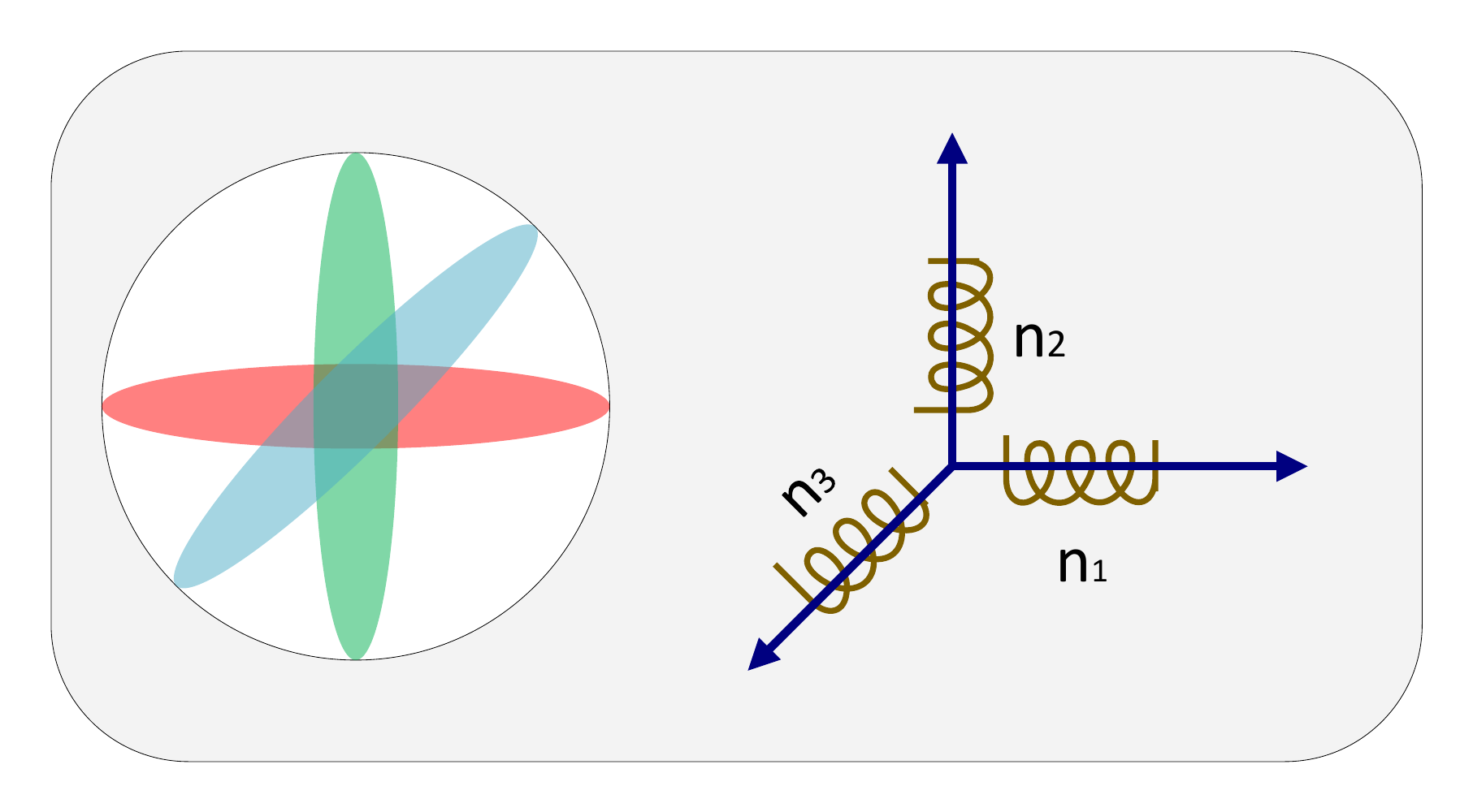}  
\caption{Tri-directional MI coils.\label{fig:trimicoil}} 
\vspace{-1.5 em}
\end{figure}

{    
MI transceivers consist of induction coils which produce quasi-static magnetic fields that can be sensed by the nearby coil. Moreover, each coil is connected to a capacitor such that the coil operates at a resonance frequency. Unlike the small commercial near field communications (NFC) coils, moderate size of coils are preferable for IoUT to cover long transmission ranges (10-100 m).
}A time-varying magnetic field is used in MI-based IoUT for communication between a transmitting and a receiving node. The coil antenna of the transmitting node produces a time-varying magnetic field which induces the current at the coil antenna of the receiver. The design of a conventional MI-based transceiver is shown in Fig.~\ref{fig:micoils} where $d_{ij}$ is the distance between the coil $i$ and $j$ , and $r_{t_{i}}$, $r_{r_{j}}$ are their radiuses, respectively. The transmitting current $I = I_0 e^{-j\omega t}$ with direct current  $I_0$  and angular frequency $\omega$  induces current in the nearby coil. However, if the transmitting and receiving coils are not well coupled, a single coil may not guarantee communication, and therefore a tri-directional coil structure was proposed in \cite{Tan2015} for efficient MI communication (see Fig \ref{fig:trimicoil}). Based on the MI phenomena, the link budget for MI-communication at high frequency and with a large number of turns in the transmitter coil $N_t$ is given in \cite{Akyildiz2006, Sun2009, Sun2010, Tan2015, Yan2018} as
\begin{equation}\label{eq: pr}
P_{r_j} = \frac{\omega \mu P_{t_i} N_{r_j} r_{t_i}^3 r_{r_j}^3\sin^2\alpha_{ij}}{16 R_0 d_{ij}^6},
\end{equation}
where $\mu$ represents the permeability of the soil, $P_{t_i}$ is the transmit power, $N_{r_j}$ is the number of turns in the receiver coil, $\alpha_{ij}$ is the angle between the axes of the two coils, and $R_0$ is the resistance of a unit length loop. {     Note that the path loss exponent for MI-based underground is  channel is much higher than the free space path loss model.}
The authors in \cite{Tan2015} experimentally validated the link budget expression in \eqref{eq: pr}. 
Based on the above channel model, different modulation schemes, such as BPSK, QPSK, and QAM were proposed in \cite{Kisseleff2014} for MI-based underground communications. Consequently, a square wave with pulse width modulation was used in \cite{Achyuta2016} to provide inductive power transfer and average data rate communication for MI-based IoUT. {     The expression for the magnetic field generated by a coil at distance $d$ was given in \cite{Achyuta2016} as
\begin{equation}
B = \frac{\mu_0 I r^2 n N_t}{2(d^2+r^2)^{\frac{3}{2}}}
\end{equation}
where $\mu_0$ is the permeability of free space,  $r$ is the radius of the coil, $I$ is the current, and $N_t$ is the number of turns of the transmitting coil. At the receiver coil, the electromagnetic field is produced as
\begin{equation}\label{E}
E = N_r A \omega B \cos\omega t
\end{equation}
where $N_r$ is the number of turns of the receiving coil, $A$ is the area of the coil, and $t$ is the instantaneous time. Based on the electromagnetic field, the voltage $V$ is measured as follows
\begin{equation}\label{voltage}
V = \frac{E R_{l_2}}{R_{l_2}+j \omega L +Z_r}
\end{equation}
 \begin{figure}
\centering
\includegraphics[width=1\columnwidth]{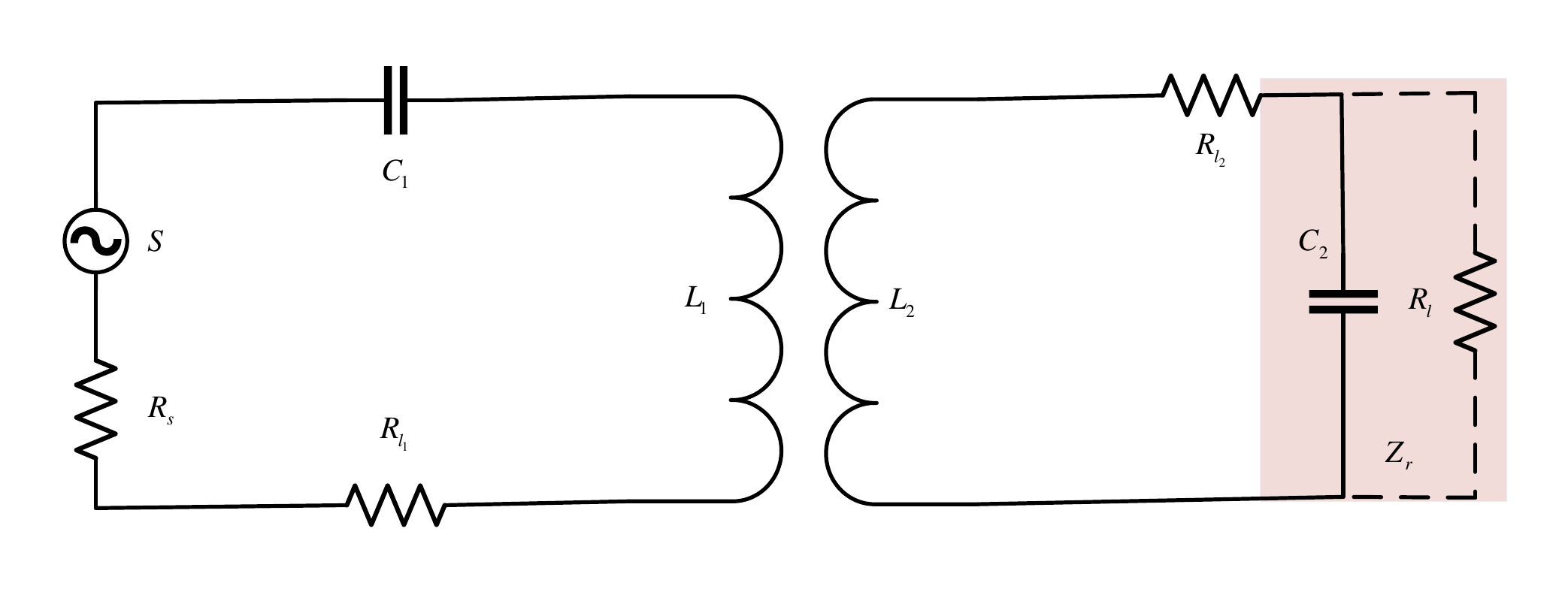}  
\caption{Equivalent circuit model for MI-based transmitter and receiver.\label{fig:micicuit}} 
\vspace{-1.5 em}
\end{figure}
where $R_{l_2}$ is the resistance of the receiving coil, $L$ is the inductance, and $Z_r$ is the load resistance by connecting the parallel capacitance resistance and pre-amplifier circuit (see Fig.~\ref{fig:micicuit}), i.e,
\begin{equation}
Z_r = \frac{R_l}{1+j \omega C_2 R_l}.
\end{equation}
Substituting \eqref{E} into \eqref{voltage} yields the expression for the measured voltage as follows \cite{Jing2014}
\begin{equation}
V =  \frac{ N_r A \omega B R_{l_2}}{R_{l_2}+j \omega L +Z_r}
\end{equation}
Based on the above forumation, the authors in \cite{Jing2014} obtained the following received  power expression
\begin{equation}\label{eq:powersci}
P_r = \frac{V_0^2 k^2 F_1^2 F_2}{2 R_{l_1} F_{1_t}}
\end{equation}
where $V_0$ is the voltage source, $k = \frac{M}{\sqrt{L_1 L_2}}$, $F_1 = \frac{\omega L_1}{R_{l_1}+R_s}$, $F_2 = \frac{\omega L_2}{R_{l_2}}$ are the quality factors of transmitting and receiving coils, respectively, $F_{1_t}$ is the instantaneous quality factor, $M = \frac{\mu_0 \pi N_t N_r r^4}{2 (\sqrt{d^2 + r^2})^3}$ is the mutual induction, and $R_s$ is the source resistance.
Authors in \cite{Jing2014} were also able to determine the size of the antenna and the number of turns in the coil for MI-based low power and low-frequency underground communications. Furthermore, they studied the impact of soil conductivity $\sigma$, on the MI-based underground communication links \cite{Ma2015}. Adding $\sigma$ to the channel, modifies the mutual inductance $M$ as follows
\begin{equation}
M_\sigma = \frac{\mu_0 \pi N_t N_r r^4}{2 \pi d^3} \exp^{-\alpha d},
\end{equation}
where $\alpha = \frac{1}{\sqrt{\pi f \mu_0 \sigma}}$ is the attenuation constant.
In \cite{Silva2016}, the soil path attenuation model with the best operating frequency range identification was presented for MI-based IoUT. Also, multiple parallel receiver circuits were used in \cite{Silva2016} to achieve higher voltage gain.
A pulse power method (use of relay coils) was used in \cite{Zungeru2016} to improve the transmission range of underground MI-based communications. The received power expression in \cite{Zungeru2016} is derived from \eqref{eq:powersci} and is given as
\begin{equation}
P_r = \frac{P_t \mu_0 N_t N_r r_t^3 r_r^3}{32 \pi d^6 C_1 R_0^2}
\end{equation}
Recently, the authors in \cite{Guo2017} suggested meta-material based MI coils for long-range subsurface communications. It was shown in  \cite{Guo2017} that using of meta-material coil provide better capacity for the same transmission distance.} 
Authors in \cite{Martins2017} and \cite{Alshehri2018}  evaluated the performance of IoUT for sandy and stone type of media where it was shown that the receiver sensitivity should be -70 dBm. Table \ref{Tablemichannel} summarizes the literature on the various physical layer issues of MI-based IoUT.

\begin{table*}
\footnotesize
\centering
\caption{{    Various physical layer issues addressed in the literature for MI-based IoUT.}}
\label{Tablemichannel}
\begin{tabular}{|p{2.0cm}|p{2.5cm}|p{7.0cm}|p{4.0cm}|}
\hline
\hline
\textbf{Ref.}           & \textbf{Frequency}& \textbf{Issue addressed} & \textbf{Design aspect}\\ \hline 
{    \cite{Sun2009}}& {    300 and 900 MHz }&  {    Underground channel modeling for MI} & {    Channel modeling}\\

{    \cite{Tan2015}}& {    0.02 and 30 MHz }&  {    Use of tri-directional MI coils for omni-directional coverage and waveguides to improve the transmission range} & {    Channel modeling and test-bed development}\\

{    \cite{Akyildiz2006}}& {    - }&  {    Discuss various issues for underground MI-based communication} & {    Cross-layer solutions}\\

{    \cite{Sun2010}}& {    10 and 300 MHz }&  {    Path loss and bandwidth analysis for underground MI communications} & {    Channel modeling}\\

{    \cite{Yan2018}}& {    100 {   kHz} }&  {    Path loss and capacity measurement for underground MI link} & {    Channel modeling}\\

{    \cite{Kisseleff2014}}& {    - }&  {    BPSK, QPSK, and QAM for the underground MI links} & {    Modulation schemes}\\

{    \cite{Achyuta2016}}& {     246 {   kHz} }&  {    Use of pulse width modulation for underground MI-based communication} & {    Testbed development}\\

{    \cite{Jing2014}}& {     246 {   kHz} }&  {    Link budget calculation for underground MI link} & {    Path loss modeling}\\

{    \cite{Ma2015}}& {     5 {   kHz} }&  {     Impact of soil conductivity on  the underground MI link} & {    Channel modeling}\\

{    \cite{Silva2016}}& {     75 {   kHz} - 30 MHz }&  {     Soil path attenuation model and best frequency selection} & {    Channel modeling and testbed development}\\

{    \cite{Zungeru2016}}& {     300 - 900 MHz }&  {     Improving transmission range by using relays and achieving higher voltage gain with  multiple parallel receiver circuits} & {     MI-based multi-hop underground communication}\\

{    \cite{Guo2017}}& {     20 - 50 MHz }&  {     Meta-material for coil design to improve transmission range and capacity} & {     Coil design}\\

{    \cite{Martins2017, Alshehri2018}}& {     10 MHz }&  {     To study the impact of different medium on the MI link} & {     Testbed development}\\
\hline 
\hline
\hline        
\end{tabular}
\end{table*}

\subsection{Networking}

The typical  MI-based IoUT network consists of buried sensors (underground things) and aboveground equipment (surface things) as shown in Fig. \ref{umi}. Hydraulic fracturing is used to inject the underground things into the well bottom or the reservoir \cite{Guo2014, Akkas2017}. The surface things can provide extended MI communication link by using large dipole antennas and large transmission power \cite{Guo2014}. Therefore, the downlink communication channel is assumed to be single hop while the up-link communication channel is multi-hop due to the limited transmission range of underground things \cite{Sun2013m}. 
The surface things can also work as anchors for the localization purpose.

Although MI-based techniques address the issue of dynamic underground channel model, the transmission distance is lower for practical use. In practical applications, the transmission distance of MI-based IoUT is improved by using the relay coils \cite{Sun2013, Kulkarni2018, Swathi2018, Pathak2018}. {     In \cite{Sun2013}, optimal deployment strategies for both one-dimensional and two-dimensional MI-based waveguide was investigated. Additionally, optimal number of relay coils were calculated to minimize the deployment cost. Minimum spanning tree (MST) algorithm was used to minimize the number of relay coils. However, MST algorithm is not robust to the coil displacement and node failure. Therefore, Voronoi-Fermat (VF) algorithm was proposed to improve the robustness without increasing the deployment cost.
Besides the use of relay coils, the authors in  \cite{Kulkarni2018} suggested the use of superconductor and meta-materials, for the MI coil design to improve the transmission range. Authors in \cite{Swathi2018} used relays to reduce the path loss for MI-based underground communication. Similarly, meta-matrial shell was used  for  MI-based transceivers  to improve their received power \cite{Pathak2018}.

\begin{figure}[t]
  \centering
  \includegraphics[width=0.5\textwidth,height=0.28\textheight]{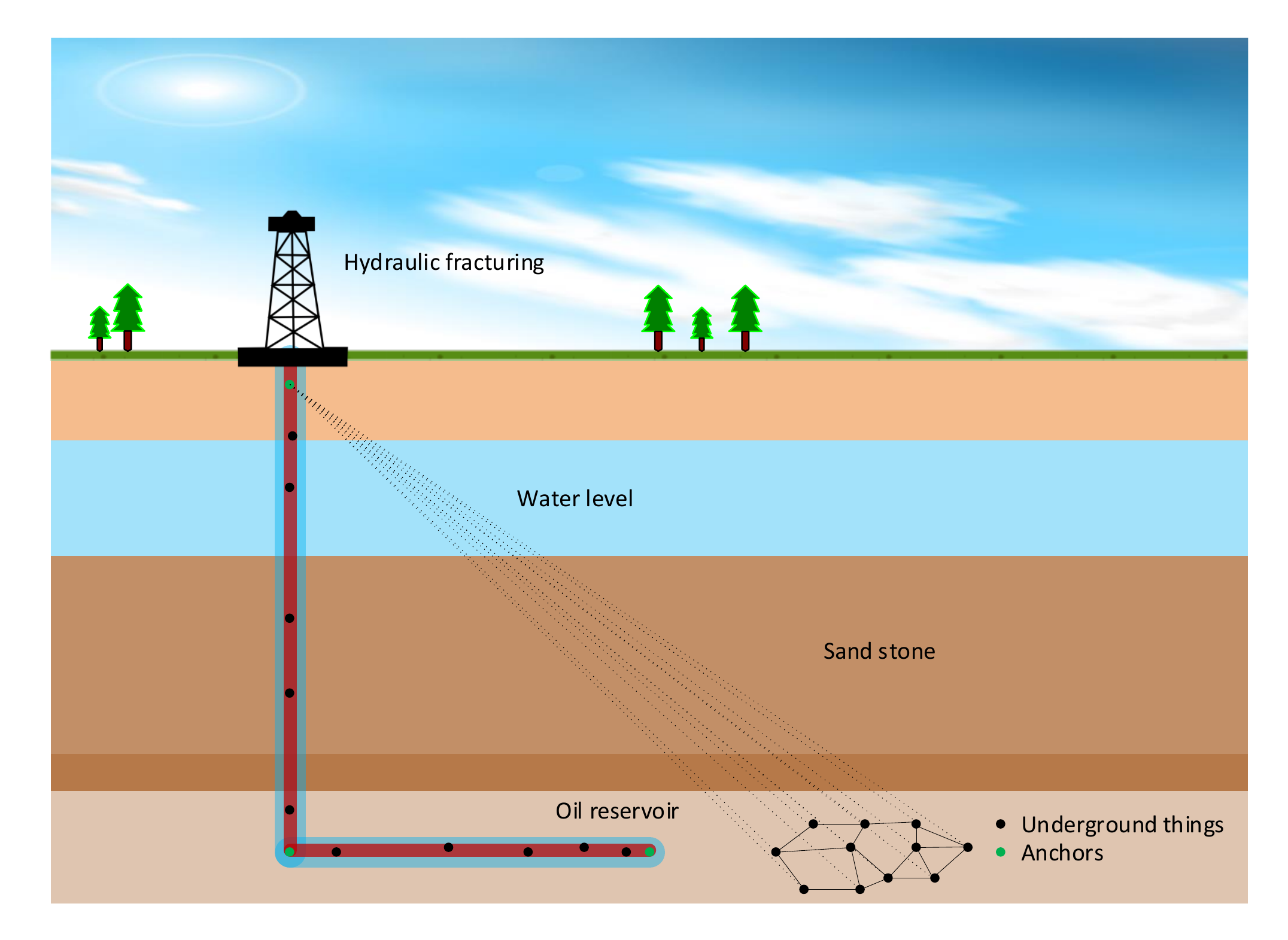}\\
  \caption{Network model for MI-based IoUT for oil and gas reservoirs.}
  \label{umi}
\end{figure}

In \cite{Kisseleff2013, Kisseleff2013b, Kisselef2015}, the network capacity of a multi-hop MI-based IoUT was evaluated. Throughput of the bottleneck link which is related to the overall network capacity was investigated in \cite{Kisseleff2013}. The polarization of MI coils was exploited to reduce the interference and improve the network capacity. The channel capacity for any arbitrary link $i$ was given in \cite{Kisseleff2013} as
\begin{equation}
C_i = \int_{-\infty}^{+\infty} \log_2 \left(1+ \frac{P_{t_i}}{L_i \cdot \text{E}\{P_n\}}\right)df,
\end{equation}
where $P_{t_i}$ is the transmit power, $L_i$ is the path loss, and $\text{E}\{P_n\}$ is the total noise power spectral density. It was shown that the proper orientation of the coils significantly reduces the interference and improve the network capacity. Similarly, optimal system parameters and topology were proposed in \cite{Kisseleff2013b} to avoid the bottlenecks in the network and to achieve higher throughput by using MI-based waveguides. Based on the above channel capacity, the throughput of link $i$ was given in \cite{Kisseleff2013b} as
\begin{equation}
T_i = \frac{C_i}{N_{d_i}(1+N_{int_i})},
\end{equation}
where  $N_{d_i}$ is the number of data streams and $N_{int_i}$ is the number of interfering nodes. The results in \cite{Kisseleff2013} and \cite{Kisseleff2013b} suggest that the use of relays and optimizing the orientation of the coils improve the throughput of a multi-hop MI-based IoUT network. Nevertheless, digital transmission schemes were introduced in \cite{Kisselef2015} for both direct and multi-hop MI-based underground links.  Frequency division multiplexing was suggested to equalize the long channel impulse response. Moreover, various modulation schemes, such as BPSK and QAM were investigated for multi-hop MI links, and it was suggested that higher order modulations could be avoided by using a bandwidth expansion mechanism.

A distributed environment aware cross-layer protocol (DEAP) was proposed in \cite{Lin2015} to satisfy the quality of service (QoS) requirement, achieve higher throughput, and reduce energy consumption. For a given MI-based underground link $i-j$, the DEAP protocol maximizes the QoS, $Q_{ij}$, and minimizes the energy consumption, $E_{ij} = U \left(\frac{P_{ij}}{R_{ij}R_{c_{ij}}}+2 E_b\right)$, where $\frac{P_{ij}}{R_{ij}R_{c_{ij}}}$ is the distance-dependent energy required to transmit a single bit, and $E_b$ is the electrical circuit energy required for a single bit transmission. $R_{ij}$ is the transmission rate, $R_{c_{ij}}$ is the channel coding rate, and $U$ is the length of the packet. This is a minimization-maximization problem where energy needs to be minimized, and the QoS needs to be maximized. Therefore, a cross-layer optimization strategy was used to save the energy and improve the network throughput.
} A two-stage cross-layer protocol called Xlayer was proposed in \cite{Chun2014} for multi-hop MI-based IoUT to guarantee the QoS requirement. Xlayer protocol was able to achieve high throughput, low delay, and low energy consumption. A full-duplex meta-material enabled MI-based communication was proposed in \cite{Guo2018} to reduce the transmission delay for multi-hop underground communication.

\begin{table*}
\footnotesize
\centering
\caption{{    Various network layer issues addressed in the literature for MI-based IoUT.}}
\label{Tableminet}
\begin{tabular}{|p{2.0cm}|p{2.5cm}|p{7.0cm}|p{4.0cm}|}
\hline
\hline
\textbf{Ref.}           & \textbf{Frequency}& \textbf{Issue addressed} & \textbf{Design aspect}\\ \hline 
{    \cite{Sun2013}}& {    10 MHz }&  {    Improvement  of the transmission range and robustness, and selecting optimal number of relays} & {    Deployment strategies}\\

{    \cite{Kulkarni2018}}& {     - }&  {    Improvement of the transmission range by using relays and meta-materials} & {     Multi-hop networking/Hardware design}\\

{    \cite{Swathi2018}}& {    300 - 900 MHz }&  {    Improvement  of the transmission range by using relays} & {    Deployment strategies}\\

{    \cite{Pathak2018}}& {    10 {   kHz} }&  {    Use of meta-material shell for the transceiver design to improve the received power} & {    Transceiver design}\\

{    \cite{Kisseleff2013}}& {     - }&  {    Investigating the effect of coil orientation and polarization on the channel capacity} & {    Interference minimization}\\

{    \cite{ Kisseleff2013b}}& {     2 and 2.5 MHz }&  {    Throughput optimization } & {    Multi-hop networking and interference minimization}\\

{    \cite{Kisselef2015}}& {     - }&  {    Optimization of system parameters for multi-hop underground MI links } & {     Maximizing the data rate }\\

{    \cite{Lin2015}}& {     7 MHz }&  {     Improving throughput, reducing energy consumption and time delay } & {     Cross layer protocol }\\

{    \cite{Chun2014}}& {     - }&  {     Throughput, delay, and energy consumption analysis } & {     Cross layer protocol }\\

{    \cite{Guo2018}}& {     10 MHz }&  {     Transmission range enhancement by using meta-material based relay coils } & {     Transceiver design}\\

{    \cite{Trang2018}}& {     300 - 1300 MHz }&  {     Connectivity analysis of multi-hop MI-based IoUT} & {     Transceiver design}\\
\hline 
\hline
\hline        
\end{tabular}
\end{table*}  
  

Recently, connectivity analysis for IoUT was provided in \cite{Trang2018} where the probability of a connected network increases with an increase in the number of underground things and low volumetric water content.{    Additionally, the connectivity performance of the EM and MI-based IoUT network was compared where the results have shown that for low volumetric water content (VWC) of the soil (1\%), the connectivity of EM and MI-based IoUT is similar. However, for the same node density when the VWC of the soil is increased to 5\%, the EM-based network becomes unconnected while the MI still maintains the connectivity.} Table \ref{Tableminet} presents the literature on various network layer issues for MI-based IoUT.

\begin{table*}
\footnotesize
\centering
\caption{{    Summary of the literature on localization for MI-based IoUT.}}
\label{Tablemiloc}
\begin{tabular}{|p{2.0cm}|p{2.5cm}|p{7.0cm}|p{1.3cm}|p{2.7cm}|}
\hline
\hline
\textbf{Ref.}           & \textbf{Frequency}& \textbf{Issue addressed} & \textbf{Dimension} & \textbf{Application}\\ \hline 
{    \cite{Markham2010}}& {    130 {   kHz} }&  {    Development of  MI-based 2D underground tracking system} & {    2D} &{     Tracking of underground animals}\\

{    \cite{Markham2012}}& {    125 {   kHz} }&  {    Testbed for  MI-based 3D underground tracking} & {    3D} &{     Underground mining}\\

{    \cite{Markham2012a}}& {    125 {   kHz} }&  {     Testbed for  MI-based 3D underground tracking} & {    3D} &{     Underground rescue operations}\\

{    \cite{Abrudan2016}}& {    1 {   kHz}, 100 {   kHz}, and 10 MHz }&  {     Investigating the impact of minerals and rocks on the localization accuracy} & {    3D} &{     Underground monitoring}\\

{    \cite{Huang2016, huang2018}}& {     - }&  {     Closed form solution for the distance estimation based on MI channel} & {    3D} &{     Underground monitoring}\\

{    \cite{Lin2017}}& {     7 MHz}&  {    Using of semi-definite programming for MI-based underground localization} & {    3D} &{     Oil and Gas reservoirs monitoring}\\

{    \cite{Abrudan2016a}}& {     10 MHz }&  {     MI-based underground localization by using a single anchor node} & {    3D} &{     Underground monitoring}\\

{    \cite{Kisseleff2017l}}& {     1 MHz }&  {     Machine learning approach for MI-based underground target localization} & {    2D} &{     Underground rescue operations}\\

{    \cite{Tian2018}}& {     - }&  {     Analytic model for distance estimation in  MI-based underground communications } & {    2D} &{     Underground monitoring}\\

{    \cite{SaeedMI2019}}& {     7 and 13 MHz }&  {     Analytical expression for the achievable accuracy of MI-based underground communications   } & {    3D} &{     Oil and Gas reservoirs monitoring}\\
\hline 
\hline
\hline        
\end{tabular}
\end{table*}

\subsection{Localization}

Localization is an essential task in wireless networks which enable various location-based services. Hence, localization techniques for the terrestrial and underwater wireless communication networks are well investigated in the past. For example in \cite{Paul2017} the authors reviewed various localization techniques for terrestrial wireless networks. Similarly, in  \cite{Saeedsurvey2018} localization techniques for marine networks are studied. The localization techniques can be classified based on the ranging technique (range-based/range-free), type of computation (centralized/distributed), and space (2D/3D). However, the literature on localization techniques for the underground wireless networks is limited due to various challenges, such as harsh and light-less underground environment, non-availability of global positioning system (GPS) signals, high attenuation, and narrow operational area. Although efforts have been made in the past to develop localization techniques for harsh environments, such as indoor and underwater, however, the underground environment does not support the use of these communication technologies, and therefore the localization techniques for the indoor and marine environment cannot be directly applied to the underground case \cite{Kisseleff2018}.

Accordingly, a two-dimensional (2D) localization technique was developed by Andrew \textit{et. al} in \cite{Markham2010} by using magnetic-induction to track animals underground. Furthermore, they extended their 2D tracking system to a three-dimensional (3D) one in \cite{Markham2012}. Moreover, they used the 3D MI-based tracking model for underground rescue operations \cite{Markham2012a}. As the propagation of MI signals is highly affected by the soil medium, therefore, the impact of minerals and rocks on the localization accuracy was studied in \cite{Abrudan2016}. It was observed in \cite{Abrudan2016} that the skin effect of the underground medium is almost negligible at very low frequencies whereas the localization accuracy depends on the attenuation properties of various underground materials. Recently, simulated annealing wasn used for MI-based terrestrial and underground wireless networks in \cite{Huang2016, huang2018} which can achieve sub-meter level accuracy.

Furthermore, a modified semidefinite programming-based relaxation technique was used in \cite{Lin2017} to determine the position of the underground sensors. A single anchor was used in \cite{Abrudan2016a} to find the location of all other underground sensors in 3D for MI-based IoUT. Trilateration, machine learning, and hybrid passive localization techniques were used in \cite{Kisseleff2017l} to estimate the position of a target node in 2D MI-based IoUT. Recently, an analytical model has been presented in \cite{Tian2018} for distance estimation in MI-based IoUT. {      The achievable accuracy of localization techniques is characterized by estimation bounds, such as the Cramer Rao lower bound (CRLB). In the past, CRLB have been derived for various wireless networks, such as internet of things (IoT) \cite{Imtiaz2016}, vehicular ad-hoc networks \cite{Ansari2018}, source localization \cite{Duan2015}, radar tracking \cite{Dersan2002}, cognitive radio networks \cite{saeed2015robust, saeed64cluster, saeed2017energy}, and underwater wireless networks \cite{saeed2018underwater, saeed2018robust, saeed2018performance}.
Therefore, in \cite{SaeedMI2019}, we derived the expression of the CRLB for the MI-based IoUT localization. The derived bound in \cite{SaeedMI2019} takes into account the channel and network parameters of MI-based IoUT.}
{     We consider a 3D oil and gas reservoir setup where the MI coils are injected to the reservoir by using hydraulic fracturing. We assume that the coils are uniformly distributed in a $15 \times 15~\text{m}^2$ fracturing area and the depth of fracture is $1.8$ km. Additionally, the buried smart objects are able to communicate with the anchors on the surface in a multi-hop fashion \cite{Lin2017}. Table \ref{Tableparameters} presents the simulation parameters which are mainly taken from \cite{Lin2017}. \begin{table}[htb!]
\footnotesize
\centering
\caption{{    Simulation parameters}}
\label{Tableparameters}
\begin{tabular}{| p{4.5cm} | p{3.5cm}|}
\hline
\hline
 Parameters              & Values     \\ \hline
 Frequency   &  $13$ MHz         \\
 Area of the Fracture            & $15$ $\times$ 15 $\text{m}^2$ \\
 Depth   & $1.8$ {   km} \\
 Radius of the coils                   & $0.01-0.04$ m \\
 Number of turns in coils          & $10-30$               \\
 Transmit power                    & $100-200$ mW                       \\
 Unit length resistance of antenna               &  $0.01$ $\Omega/m$         \\
 Temperature            & $418$ K              \\
 Noise variance                 & $0.1-0.8$ m\\
 Number of underground things & $60$\\
\hline
\hline        
\end{tabular}
\end{table}
}


\begin{figure}
\centering
\includegraphics[width=0.8\columnwidth]{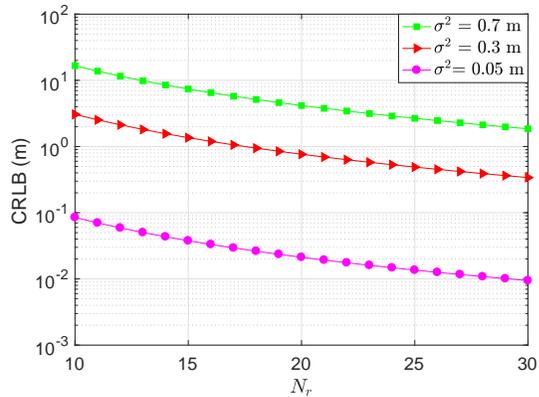}  
\caption{CRLB vs. Number of turns in the receiver.\label{fig:turns}} 
\end{figure}

\begin{figure}
\centering
\includegraphics[width=0.99\columnwidth]{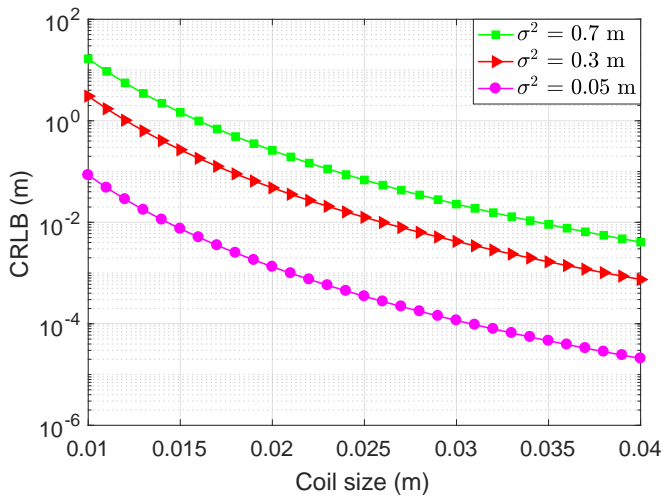}  
\caption{CRLB vs. Coil size.\label{fig:coilsize}} 
\end{figure}

\begin{figure}
\centering
\includegraphics[width=0.99\columnwidth]{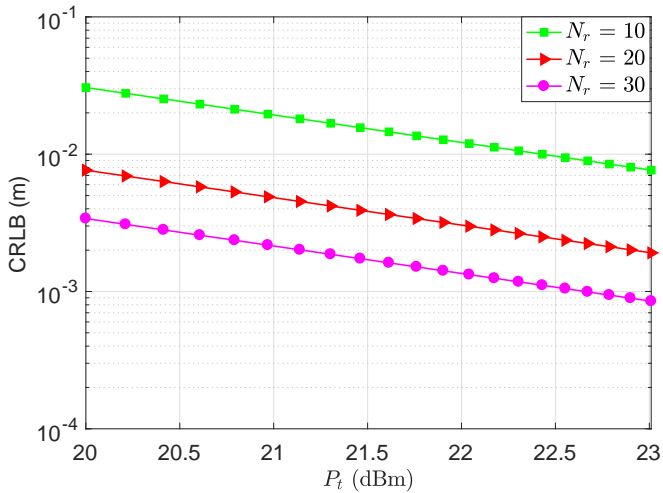}  
\caption{CRLB vs. Transmit power.\label{fig:power}} 
\vspace{-1.5 em}
\end{figure}


Fig. \ref{fig:turns} shows the impact of the number of turns in the MI coil and the noise variance on the achievable accuracy. The values of the noise variance are kept as 0.05, 0.3, and 0.7 m, respectively whereas the frequency is 13 MHz. Fig. \ref{fig:turns} suggests that to get better localization accuracy more number of turns in the coil are required however increasing the number of turns may increase the size of the coil. However, the harsh underground environment requires a small size of the MI coil. Hence, the impact of coil size is examined in Fig. \ref{fig:coilsize} which suggests that increasing the size of the coil improve the accuracy. Thus, there is a trade-off between the size of the coil and the localization accuracy which should be taken into account before the deployment.

Moreover, we have also tested the impact of transmission power on the achievable localization accuracy in Fig. \ref{fig:power}. Commercially available MI coils have transmission power in the range of 100 – 200 mW (20-23 dBm). Thus we kept the transmit power in the range of 20-23 dBm with the variable size of the coils. Fig. \ref{fig:power} shows that with the increase in transmit power the achievable accuracy improves. Accordingly, the above results suggest that the achievable accuracy of any localization algorithm for MI-based IoUT is the function of the number of turns in the MI coil, noise variance, size of the coil, transmit power, frequency, and the number of anchors. Therefore, all these parameters are important to design a robust and accurate localization technique for IoUT.

\subsection{Charging of the MI Coils}
Lifetime is an important parameter for IoUT due to the harsh underground environment. Therefore, research efforts have been made to improve the lifetime of the IoUT \cite{Kisseleff2016}. A charging method for IoUT has been proposed in \cite{Kisseleff2016} where a virtual magnetic relay network and optimized routing protocol was used to reduce the energy consumption. However, the charging efficiency of the proposed system in \cite{Kisseleff2016} remains very low for even moderate size of coils. Consequently, in \cite{Alshehri2017} an optimized energy model framework was proposed for linear topology.
The problem of charging underground coils is still open research problem and other options, such as energy harvesting can be investigated.
\begin{figure}[t]
  \centering
  \includegraphics[width=0.3\textwidth,height=0.4\textheight]{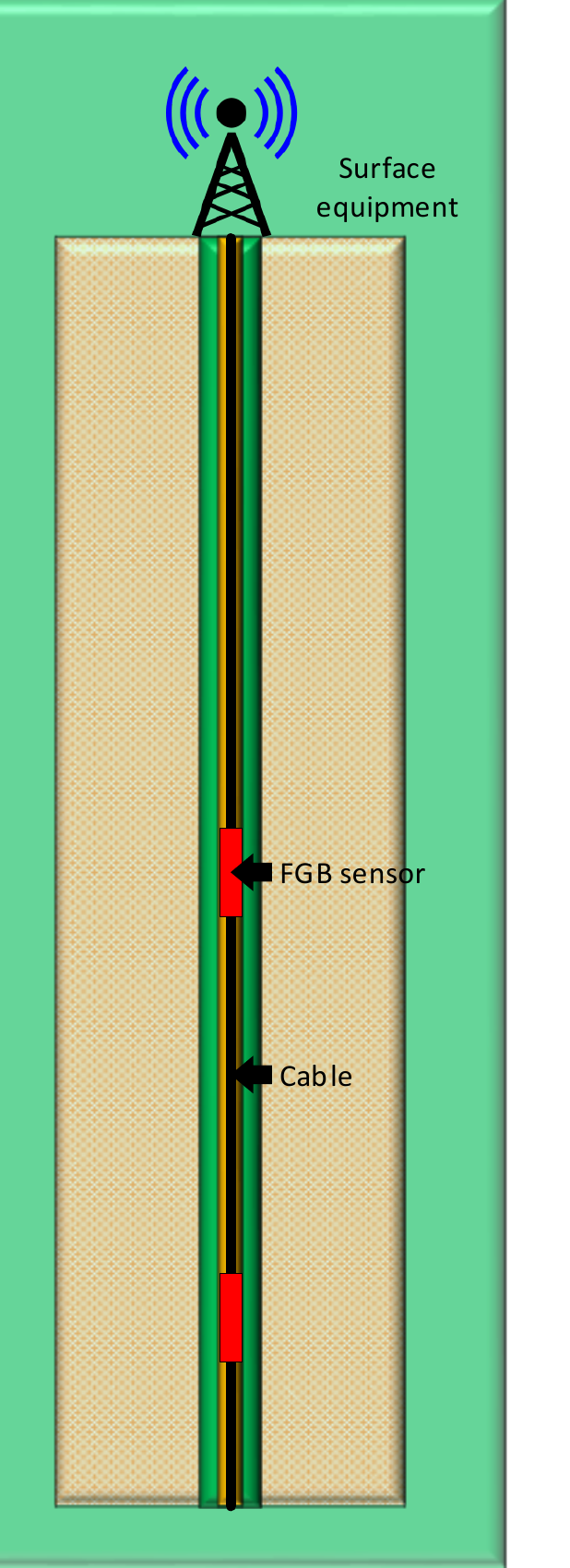}\\
  \caption{Fiber optic monitoring system for IoUT.}
  \label{opnet}
\end{figure}

\section{Wired Communications for IoUT}
Wireless communication channels reduce the complexity and cost of underground monitoring. However, existing wireless technologies fail to provide timely, reliable and accurate solutions, especially for the deep underground monitoring. Hence, wired technologies, such as coaxial cable and optical fiber are used for down-hole monitoring \cite{Schroeder2002, hernandez2008high,algeroy2010permanent, mijarez2013communication,BALDWIN2018}.  {    In 2003, Intelliserv provided a high speed down-hole telemetry system based on coaxial cable as an alternative to the low speed MPT and electromagnetic systems \cite{Intelli}. This system provided real time bi-directional communication at the data rate of 57,000 bits per second which has greatly enhanced the bandwidth for down-hole communication. Coaxial-based system consisted of  inductive coils and repeaters along the high strength coaxial cable \cite{hernandez2008high}. Similarly, a fiber optic-based temperate sensing system was developed in \cite{algeroy2010permanent} to monitor the oil and gas reservoirs. The authors in \cite{ mijarez2013communication} developed a down-hole communication system in the presence of high temperature and pressure. The proposed system in \cite{ mijarez2013communication} was based on the SNR characterization of the whole channel where optimal frequency for the transmission is selected.} Since for higher data rates, optical fiber has replaced co-axial based wired communication. In this section, we cover the literature on optical fiber-based communication systems for underground applications.

\begin{table*}
\footnotesize
\centering
\caption{{    Summary of the literature on wired communications for the IoUT.}}
\label{Tablemwired}
\begin{tabular}{|p{2.0cm}|p{2.1cm}|p{7.0cm}|p{2.9cm}|p{1.2cm}|}
\hline
\hline
\textbf{Ref.}           & \textbf{Type}& \textbf{Issue addressed}  & \textbf{Application}& \textbf{Year}\\ \hline 
{    \cite{Schroeder2002}}& {     Optical fiber }&  {    Study on the use of optical fiber for Oilfield industry}  &{     Oilfield monitoring}& {    2002}\\

{    \cite{hernandez2008high}}& {     Coaxial cable }&  {    Development of high speed down-hole communication system}  &{     Down-hole telemetry}& {    2008}\\

{    \cite{algeroy2010permanent}}& {     Optical fiber }&  {    Down-hole communication temperature sensing}  &{     Management of oil and gas reservoirs}& {    2010}\\

{    \cite{mijarez2013communication}}& {      Coaxial cable }&  {    Down-hole communication in the presence of high pressure and high temperature}  &{     Management of oil reservoirs}& {    2013}\\

{    \cite{BALDWIN2018}}& {     Optical fiber }&  {    Discussion on various applications of fiber optic sensing}  &{     Underground monitoring}& {    2018}\\

{    \cite{kragas2001optic}}& {     Optical fiber }&  {    Development of fiber optic based down-hole telemetry system}  &{     Down-hole monitoring}& {    2001}\\

{    \cite{kersey2000optical}}& {     Optical fiber }&  {    Review of fiber Bragg grating sensors for down-hole monitoring}&{     Down-hole monitoring} & {    2000 }\\

{    \cite{Zhang2007}}& {     Optical fiber }&  {    Field tests by using FBG-based seismic geophones} &{     Oil and Gas reservoirs monitoring}& {    2007}\\

{    \cite{zhou2012simultaneous}}& {     Optical fiber }&  {    Multiplexing of temperature and pressure FBG sensors} &{     Oil and Gas reservoirs monitoring}& {    2012 }\\

{    \cite{Prevedel2015}}& {     Optical fiber }&  {     FBG sensors-based testbed development} &{     Geophysical observations}& {    2015}\\
\hline 
\hline
\hline        
\end{tabular}
\end{table*}

Fiber optic sensing technologies have been applied to several commercial and industrial application in the past two decades. These sensing technologies can provide both sensing and information transfer in the harsh environment. Due to these attributes, they are well suited for the harsh environment in oil and gas reservoirs \cite{kragas2001optic}. {     In \cite{kragas2001optic}, the authors discussed the deployment of underground fiber optic-based  monitoring systems, including the assembly of down-hole sensors, data measurement, and installation of the sensors. The authors in  \cite{kragas2001optic} also briefly discussed about the installation of the pressure and temperature sensing system in a land well and in the Gulf of Mexico. It was concluded in \cite{kragas2001optic} that fiber Bragg grating (FBG) sensors offer additional  advantages, such as high flexibility, high stability, multi-point sensing, and their intrinsic nature to the fiber.} The fiber optic based underground sensing system consists of FBG sensors which are connected to the optical fiber cable with the help of ultraviolet photo-inscription method \cite{kersey2000optical} (see Fig. \ref{opnet}). 

{    
The authors in \cite{NELLEN2003} proposed a FBG based real-time temperature and fluid monitoring system of oil bore-holes. The FBG-based sensor converts the fluid pressure by means of a mechanical transducer into fiber optic strain. The strain of the transducer was given as  \cite{NELLEN2003}
\begin{equation}
s = \frac{s_l - \lambda (s_r + s_t)}{E},
\end{equation}
where $s_l,~s_r,~s_t$ are the longitudinal, radial, and tangential stresses, respectively, $\lambda$ is the Poisson number, and $E$ is the Young's modulus. Based on the inner and outer radii of the tube, $a_i$ and $a_o$, the above expression is re-written as
\begin{equation}
s = \frac{p a_i^2 (1-2\lambda)}{E(a_o^2-a_i^2)},
\end{equation}
where $p$ is the pressure.
In \cite{Zhang2007}, Yan \textit{et. al} designed FBG based seismic geophone for oilfield exploration which showed better sensitivity than the conventional geophones for 10-70 Hz range of frequencies. Moreover, field tests were performed  in \cite{Zhang2007} where the FBG-based seismic geophone was more immune to the EM interference, and had less non-linear distortion as compared to the conventional FBG sensors.

Wavelength division multiplexing was used in \cite{zhou2012simultaneous} to combine the information from two fibers onto a single fiber for the temperature and pressure sensing in a well-bore. The simultaneous sensing of pressure and temperature, is a cost-effective solution. Experimental results were provided  in \cite{zhou2012simultaneous} showing that the proposed multiplexed approach is stable and accurate.}
Recently, the authors in \cite{Prevedel2015} installed an optical fiber-based down-hole monitoring system at the shoreline of Marmara sea in Turkey to provide geophysical observations. In short, optical fiber based underground monitoring systems provide high-speed communication and are immune to electromagnetic interference which mainly depends on the development of fiber grating sensors \cite{Wu2019}. {    Table \ref{Tablemwired} summarizes the major contributions on wired communications for IoUT.}

\begin{figure*}[t]
  \centering
  \includegraphics[width=0.7\textwidth,height=0.13\textheight]{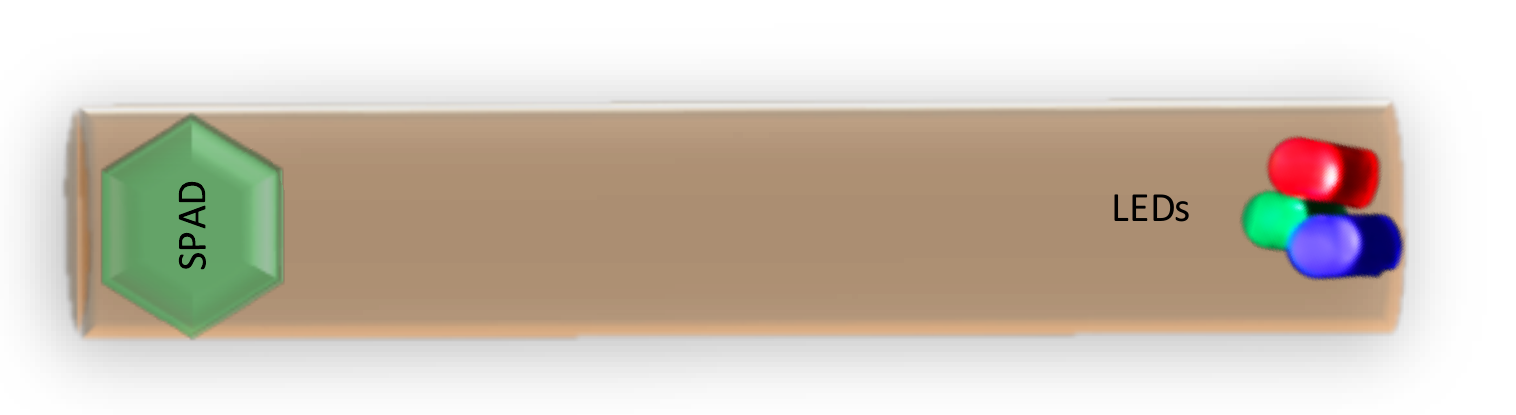}\\
  \caption{VLC based IoUT for gas pipelines.}
  \label{vlc}
\end{figure*} 

\section{Visible light communications for IoUT}
Light cannot pass through soil therefore visible light communication (VLC) can only be used for down-hole monitoring in gas fields. To the best of the author's knowledge, the only VLC-based down-hole monitoring systems were proposed in \cite{Li2014} and \cite{Miramirkhani2018}. In \cite{Li2014}, light emitting diodes (LEDs) were used at the bottom of the pipeline, and a single photon avalanche diode (SPAD) was used as a receiver at the surface (see Fig. \ref{vlc}). The proposed system in \cite{Li2014} was able to achieve the data rate of 1-5 {   kbps} for the depth of 4000 m. {     An array of SPAD receivers were set at the surface of the pipe where the number of photons received were estimated as follows:

\begin{equation}
N_p = C_e \sum_{n = 0}^R \frac{(\beta + 1) A P_t}{2 \pi d^2 E_p} \cos^\beta(\phi(n))C_p,
\end{equation}
where $C_e$ is the photon detection efficiency, $A$ is the area of the SPAD, $d$ is the distance, $\phi$ is the angle of radiation, $P_t$ is the average power of LED, $E_p$ is the energy per photon, $R$ is the total number of reflections, $C_p$ is the reflectivity of pipe, and $\beta = \frac{- \ln(2) }{\ln (\cos(\theta))}$, where $\theta$ is the half power beamwidth. Modulation scheme of non-return-to-zero on-off keying (NRZ-OOK) was used where binary ``1" is represented by a high positive voltage and binary ``0" is represented by a low voltage. Poisson distribution was used in \cite{Li2014} to model the noise where the cumulative distribution function (CDF) is given as follows:
\begin{equation}
P_c(x, \lambda) = \exp^{-\lambda}\sum_{i=0}^x \frac{\lambda^i}{i ! },
\end{equation}
where $\lambda$ is the mean value of the distribution. The probability of error is calculated from the above CDF as follows:
\begin{equation}
P_E = \frac{1}{2}[1-P_c(N_t,N_0)+P_c(N_t,N_1)],
\end{equation}
where $N_t = \frac{N_0 + N_1}{2}$ is the threshold value, $N_0$ is the average number of photons when binary ``0" is transmitted, and $N_1$ is the average number of photons when binary ``1" is transmitted. It was concluded  in \cite{Li2014} that the BER improves with reducing the half power beamwidth angle and increasing the LED's power.
}
Nevertheless, the authors in \cite{Li2014} assume empty pipeline which is not realistic.  Hence,  in \cite{Miramirkhani2018} the authors provided a channel model for VLC-based underground gas pipelines in the presence of methane gas. {    Ray tracing technique was used to model the channel. The channel impulse response (CIR) for the ray tracing technique is expressed as \cite{Miramirkhani2018}
\begin{equation}
h(t) = \sum_{i=1}^{N_r} P_i \delta(t-\tau_i)
\end{equation}
where $N_r$ is the number of rays, $P_i$ is the power from $i$-th ray, $\tau_i$ is the propagation time of $i$-th ray, and $\delta(\cdot)$ is Dirac delta function. By taking the Fourier transform of the CIR yields the frequency response of the optical channel as follows:
\begin{equation}
H(f) = \int \sum_{i=1}^{N_r} P_i \delta(t-\tau_i) \exp^{-j2\pi f t}dt
\end{equation}
Based on $H(f)$, the frequency response is commonly modeled as
\begin{equation}
H_{L}(f) = \frac{1}{1+\frac{jf}{f_c}},
\end{equation}
where $f_c$ is the cutt-off frequency. By taking into account the characteristics of LED, the effective channel response in frequency domain is expressed as $H_e(f) = H_{L}(f)H(f)$. Based on the expression of the channel response, the BER is given as
\begin{equation}
\text{BER} = \frac{2(M-1)}{M \log_2(M)} Q\left( \frac{1}{M-1}\sqrt{\frac{(P_t h_e r_s)^2 T_s}{N_o}}\right),
\end{equation}
where $M$ is the constellation size for $M$-ary PAM schemes, $h_e(t) = F^{-1}(H_e(f))$, $P_t$ is the average optical power transmitted, $r_s$ is the responsivity of photodetector, $T_s$ is the sampling interval, $N_o$ is noise power spectral density, {   and $F^{-1}(\cdot)$ is the inverse Fourier transform.}}
Different pulse amplitude modulation (PAM) schemes were tested, and the results have shown that for a target BER of $10^{-6}$, 8-PAM can reach the target distance of 22 m. Similarly, higher order PAM provides a better data rate but reduced transmission range with a minimum of 3.82 m for 512-PAM. Nevertheless, the research on VLC-based IoUT is in its infancy and need to be examined in the future.

\section{Future Research Challenges}
The recent advances in IoUT have broadened the scope of this research area. Hence, in this section, we provide various new challenges for the IoUT. Table \ref{Tablechal} presents the significance of each research challenge for a specific underground application.

\begin{table}
\footnotesize
\centering
\caption{Significance of each research challenge for various applications.}
\label{Tablechal}
\begin{tabular}{|p{2.5cm}|p{1.4cm}|p{1.7cm}|p{1.0cm}|}
\hline
\hline
\textbf{Research Challenge}           & \textbf{Agriculture} & \textbf{Seismic exploration} &  \textbf{Oil \& Gas}\\ \hline
\textbf{Deployment} &  	Medium &High &High\\ \hline
\textbf{Channel modeling} & Medium &Medium & High\\ \hline
\textbf{Transmission range} & Low &High & Medium\\ \hline
\textbf{Latency} & Low & Low & Medium\\ \hline
\textbf{Reliability} & Low & Medium & High\\ \hline
\textbf{Security} & Medium & High & High\\ \hline
\textbf{Scalability} & Low & Medium & Medium\\ \hline
\textbf{Robustness} & Low & Medium & High\\ \hline
\textbf{Networking} & High & Medium & Medium\\ \hline
\textbf{Cloud computing} & High & Medium & Low\\ \hline
\textbf{Fog computing} & Low & Medium & High\\ \hline
\textbf{Localization} & Medium & High & Medium\\ \hline
 
\hline
\hline        
\end{tabular}
\end{table}

\subsection{Deployment}
The deployment of smart objects for IoUT in the harsh underground environment is a challenging issue \cite{Kisseleff2018}. Installation and management of smart objects underground are much more difficult compared to the terrestrial networks. Moreover, the underground objects can be easily damaged during the digging process. Hence, efficient deployment of the smart objects is required to minimize the installation cost of the IoUT.
 For example, a smart object with high energy requirement should be deployed near the surface for ease of management as the replacement of batteries in the underground environment is challenging. Moreover, to avoid the replacement of batteries, a battery with high capacity should be used, and power saving protocols should be applied. {     Deployment of underground objects is more challenging in case of seismic exploration and underground drilling as compared to the agriculture applications due to high depth. Hence, various efforts are made in the past based on different metrics and parameters for the optimal deployment of underground objects. For example, in MI-based IoUT, the orientations and polarization of the coils are managed to reduce the power reflections \cite{Sun2013}. Similarly, horizontal and vertical deployment strategies were introduced in \cite{Kisseleff2013} to reduce the complexity of the network. Moreover, Voronoi tessellation was used in \cite{Kisseleff2013b} to select the underground objects to be connected to the relay network. Besides, the effect of heterogeneous soil on the path loss also needs to be taken into account for the optimal deployment of underground objects.}
Unfortunately, the research work on the efficient deployment of IoUT is limited, which takes into account the various system and network parameters, such as deployment depth, number of smart objects, lifetime, and routing.

\subsection{Channel Modeling}  
In terrestrial communications, the strength of the EM signal decays with the square of the distance while in the soil, the decay is much faster due to attenuation from the soil medium \cite{Silva2009}. The major loss factors for a given frequency in the soil are the permittivity and conductivity of the type of soil. Hence, MI-based communication was introduced for the IoUT which is based on magnetic field propagation \cite{Tan2015}. {     MI-based underground communications is also affected highly by the heterogeneous soil medium. However, most of the existing works assume homogeneous soil medium which is not realistic. In heterogeneous soil medium, the magnetic fields are attenuated differently at each layer of the soil. Therefore, the authors in \cite{Wait1971} defined different scaling factors for various depths. Similarly, the signal propagation through the heterogenous medium was analyzed in \cite{Kisseleff2015} where the equivalent skin depth was calculated for each layer of soil. The path loss for a single link in MI-based underground communications is characterized in \cite{Sun2010}. Moreover, asymmetric transceivers was introduced in \cite{Guo2017} to provide long range underground communications in case of misalignment between the MI coils.}
Although the path loss for each type of communication channel has been extensively analyzed in the past, unfortunately, few efforts have been made to provide a fully functional wireless solution for IoUT with practical signal transmission schemes to verify the channel models. Nevertheless, the demanding applications of the underground stimulate the research in this direction.

\subsection{Limited Transmission Range}
The MI technology has certain advantages, such as prone to multi-path fading and boundary effects which makes it ideal for the communication and localization in the underground environment \cite{Tan2015}. However, MI technology also has some disadvantages; the most significant one is the limited transmission range due to high path loss in the soil. {     Few efforts have been made in the past to improve the transmission range for MI-based underground communications. For example, the use of relay coils was proposed in  \cite{Sun2013} and \cite{Kulkarni2018} to extend the transmission range. Also, the use of superconductor and meta-materials was recommended in  \cite{Kulkarni2018} and \cite{Pathak2018} to improve the transmission range of MI-coils.
Besides that, large coils with high transmission power were used for the long-range downlink communication and localization in \cite{Lin2017}, however, it may not be a practical solution. Hence, the limited transmission range of buried smart objects in IoUT is still a significant issue which is still an open research problem.}

\subsection{Low Latency and Reliable Communications}
Most of the applications of IoUT require low latency and reliable communication. For example, the underground things are deployed in the oil and gas wells to perform critical sensing tasks, such as temperature/pressure, pipe leaks and gas leaks, and therefore the data should be reliable and received at the surface with low latency. If the sensed data from even a single sensor is not received on time, it can lead to a disaster. As the performance of the IoUT is highly susceptible to the harsh underground environment, it can lead to unreliable and high latency communication. {     Although the requirement of low latency and reliable communication is essential for both terrestrial IoT and IoUT, it is important to note that the operating environment and regulations in underground are different than the conventional IoT networks. There is no single communication system which can meet all the requirements in the harsh underground medium. For example, the wired solutions can provide reliable and low latency underground communications, but they are not scalable, are complex, and have a high cost. On the other hand, the wireless solutions are scalable, have low complexity, and have low cost but are susceptible to the harsh underground environment.
Hence, a reliable and low latency architecture is required for IoUT which can count for sensor failures and minimize the transmission delay.}

\subsection{Security}
Although efforts have been made in the past to model the channel for IoUT \cite{Silva2009, Tan2015}, unfortunately, security for IoUT did not get much attention from the research community. Security of IoUT includes the security of equipment and the communication protocols. Various attacks, such as node replication, signal jamming, and wormhole can be launched to destabilize the operation of the IoUT. The security breach can also exhaust the network resources by triggering false alarms and responding to false alarms. Various security issues for a cloud-based IoT are addressed in \cite{Roman2011} where the issues of malicious attack, forward and backward security, node compromise attack, and semi-trusted cloud security are identified. Similarly, the authors in \cite{Evans2012} provided efficient data tagging technique by using information flow control (IFC) that improves the security of the data. However, tagging of resource-constrained IoT devices is expensive. 
Moreover, the authors in \cite{Garcia2013} presented a secure architecture for IoT networks based on datagram transport layer security (DTLS) and host identity protocol (HIP). A detailed survey on security issues in IoT networks can be found in \cite{Sicari2015}. All of the above works are focused on security and privacy for terrestrial IoT networks that can be adapted for IoUT by adding the constraints of the underground environment. {    For instance, the older control and information systems for oil and gas fields are transforming into a digitized IoUT network. The security mechanisms are also needed to be updated because these systems were never designed to be globally connected.  Hence, such systems can easily be targeted by the cyber-attacks, leading to a disaster. One solution for the security in IoUT networks is to investigate the blockchain technology, which can significantly reduce the possibility of cyber-attacks. The distributed ledger in blockchain can be an ideal solution to provide de-centralized security.}

\subsection{Scalability}
Routing overhead, higher network density, and node failures can cause scalability issue for IoUT. Moreover, the high energy consumption and limited memory of underground sensor nodes limit the scalability of the network. Additionally, the IoUT system may consist of sensor nodes developed by different vendors which can lead to interoperability issues. {     Several works on scalability issue exist for various types of IoT networks. For example, 
in \cite{Gharbieh2017}, the authors investigated the scalability issue in cellular networks for IoT devices by using spatiotemporal stochastic modeling.
Similarly, structure-aware and self-adaptive wireless sensor network setup was proposed in \cite{Li2007} to address the scalability issue in a tunnel. Moreover, a middleware protocol was introduced in \cite{Li2007} to cope with the interoperability issue. The proposed protocol in \cite{Vresk2016} connects heterogeneous IoT devices. The above works address the issue of scalability and interoperability for terrestrial IoT networks. 
However, the harsh underground environment should be taken into account to modify these techniques for IoUT. 
{   For example, in the case of buried smart objects, the increased path loss of wireless communication signals in soil limits the connectivity of a large-scale IoUT network. In \cite{Tooker2012}, the authors investigated the problem of scalability and lifetime maximization for agricultural IoUT where a mobile sink node was used to connect the sparsely deployed nodes, and energy harvesting was used to improve the network lifetime. Recently, a large-scale IoUT was implemented in \cite{Zhang2019}, implementing a single-hop star topology which was able to support up to one million devices. Besides a few of the above solutions, IoUT requires the development of self-healing and self-organizing techniques to overcome the scalability issue.}

\subsection{Robustness}
Robustness is another crucial issue for IoUT which has not been addressed in the past. In an underground resource-constrained environment where there are various challenges, such as dynamic topology, energy constraint, and sparsity of nodes, robustness is critical.      In \cite{Luo2015}, the authors proposed a small world model to improve the robustness and latency of heterogeneous IoT network where the local importance of the smart object was taken into account. Robustness for terrestrial IoT networks is well-studied from different perspectives. For example, form localization aspect, detailed discussion about the robustness  in terrestrail IoT can be found  in \cite{Chen2017}. The literature on robustness for underground networks is only limited to the mining application. For example, a robust mesh wireless network was proposed in \cite{Kennedy2006} for underground mining which enhances both reactive and precautionary measures in safety and emergency situations. {    Since communication in IoUT networks suffers from an extremely harsh environment, it is challenging to develop robust communication and data gathering schemes. For example, the heterogeneous soil medium and water content of the soil limit the transmission range of electromagnetic waves. However, magnetic induction is more robust to the soil medium and water contents. Nevertheless, magnetic induction suffers from low transmission range and require perfect orientation between the coils. However, the research on using magnetic induction and its robustness for underground communications is still not mature and require further investigation.}



\subsection{Hybrid Sensing}
A hybrid sensing system for IoUT integrates signals from multiple types of sensor systems for the detection and localization of an event. For example, a network of long-term underground fiber sensors can be fused with short-term ground penetrating radars for the detection and localization of an underground event. {     Another example of hybrid sensing network is SoilNet system proposed in \cite{Bogena2009}. In SoilNet, EM-waves based Zigbee network is used for above-ground communications and wired communication is used for underground links between the buried nodes. Similarly, hybrid EM and MI-based sensing network can be used where the EM can provide long-range down-link communication while short-range MI can be used to provide multi-hop uplink communication \cite{Lin2017}.
Hence,  hybrid sensing technologies and new concepts, such as crowd-sensing can proactively detect and localize underground events, and improve the efficiency of IoUT.}
\subsection{Software Defined Networking}
Although the research on IoUT is still in its infancy, researchers are already looking to develop networking solutions for the smart objects connected underground.    Unlike conventional networking, software-defined networking (SDN) provide a scalable, secure, and reliable solution. Due to these advantages of SDN, it is used for underwater communications systems where the underwater sensors use in/out-band control channels to communicate with the surface station \cite{softwater2016}. The surface station acts as an SDN controller to separate the data plane and the control plane. A similar approach can be used for IoUT, to provide an appropriate networking  \cite{puente2018software}. By employing SDN for the IoUT reduces the network complexity, improve load balancing, provide congestion control, efficiently utilize the network resources, improve the network lifetime, and reduce the latency. {   For example, in IoUT networks for oil and gas reservoirs monitoring, SDN  can provide a global network view of the underground smart objects enabling efficient management of the network. Besides, for large-scale IoUT networks in agricultural applications, SDN can provide a scalable network management solution. Moreover, data visualization tools such as multidimensional scaling \cite{Saeed2018survey} can be incorporated at the SDN controller to correlate the data from the smart objects. Hence with all these advantages, the SDN paradigm needs to be examined for the IoUT.}
\subsection{Big Data}
IoUT is going to generate a large amount of exploration data which include data coming from various applications, such as agriculture, seismic surveying, and oil/gas fields.  Hence this massive amount of data need to be organized for a proper analysis, a metric calculation, or an event correlation to make accurate decisions \cite{hajirahimovaopportunities}. There has been a lot of research efforts made recently to integrate IoT and big data for conventional IoT networks. For example, Atzori \textit{et al.} presented an overview of enabling technologies and potential applications of big data for IoT networks \cite{ATZORI20102787}. Similarly, context-aware computing and its usage for IoT was discussed in \cite{Perera2014} and the references therein. These works motivate the use of big data analytics including descriptive, predictive and prescriptive analysis for underground applications as well.
{    For example, in oil and gas IoUT networks, a large amount of exploration data is generated where managing such a massive amount of data is of major concern to the oil and gas industry. It is reported in \cite{MOHAMMADPOOR2018} that the Geoscientists spent half of their time in managing and processing the collected data. Big data can be used to handle this massive amount of data and conduct various analysis such as scheduling and drilling.  In short, proper data analytics tools need to be developed to support the large amount of data produced in IoUT.
}

\subsection{Cloud and Fog Computing}
The real-time and localized operations can be enabled for IoUT by integrating it with cloud and fog computing. Moreover, cloud/fog computing can provide various services for IoUT, such as scalability, location awareness, low latency, and mobility. In the recent past cloud computing has been used to provide maintenance for the oil and gas industries while fog computing has been used to reduce the data traffic and provide the analysis of the data at the network edge \cite{PERRONS2013732}. {    Also, analysis of a large amount of data generated in oil and gas industries is a big challenge; especially, in upstream where data-rich operations occur, e.g., real-time drilling, seismic exploration, and integrated processes. Hence, fog computing can provide localized real-time analytics of the data avoiding the communication delays and ensure a faster reaction to an event. Also, it can execute the drilling operations in a closed loop from a remote side by supervising through a cloud application. Hence, in time-critical applications by the time the data reach to the cloud for analysis, the opportunity to make a decision might be gone. Therefore, fog computing techniques should be integrated with the IoUT to support the in-time decision.}

\subsection{Robust and Accurate Localization Methods}
Localization for IoUT enables numerous applications, such as geo-tagged sensing data, monitoring of the underground environment, and optimized fracturing. Few efforts have been made in the past to find the location of buried underground objects for MI-based IoUT. {     For example, in \cite{Markham2012}, a testbed was developed based on MI communication to track the underground objects. Moreover, the impact of mineral and rocks on the localization accuracy was investigated in \cite{Abrudan2016}. Similarly, a semi-definite programming-based localization technique was developed in \cite{Lin2017} for MI-based underground sensor networks. Nevertheless, the achievable accuracy for MI-based underground networks was investigated in \cite{SaeedMI2019}. All of these works are focused on MI-based communications where the work on EM, acoustic, and VLC do not exist.
Therefore, robust and accurate three-dimensional methods need to be investigated based on each technology to enable the applications as mentioned earlier.}

\section{Summary and Conclusions}
In this article, we have surveyed various communication, networking, and localization techniques for the internet of underground things (IoUT). In section II, we have presented the literature on EM waves-based IoUT where we have briefly discussed the channel model and networking solutions. The primary applications of EM waves based IoUT are in agriculture due to their low penetration depth in the soil. In section III, acoustic waves-based solutions are presented for underground communications. Due to their low frequency, acoustic waves are mostly used for seismic exploration and down-hole communication during drilling. We have gathered the acoustic-based underground communication techniques in Table III and IV, where we have briefly stated their applications.
Nevertheless, in section IV we have presented a well-known mud pulse telemetry system which is commercially used for down-hole communication but suffers from various impediments, such as attenuation, dispersion, mud pump noise, and gas leakage. Furthermore, the most recent technology for IoUT, i.e., magnetic induction is presented in section V and various issues addressed in literature are stated in Table V and VI. Moreover, high speed wired solutions,   such as coaxial cable and optical fiber for the IoUT are presented in section VI. In section VIII, visible light communication-based IoUT are briefly discussed which did not get much attention of the researchers yet. Finally, we have presented various future research challenges which can be investigated by the researchers. In short, the research on communication, networking, and localization for IoUT still have a long way to go and require the attention of both academia and industry.
\bibliographystyle{../bib/IEEEtran}
\bibliography{../bib/IEEEabrv,../bib/nasir_ref}

\begin{IEEEbiography}[{\includegraphics[width=1in,height=1.25in]{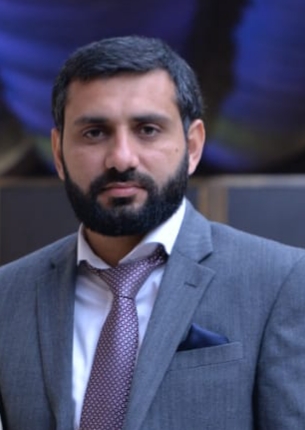}}]{Nasir Saeed}(S'14-M'16) received his Bachelors of Telecommunication degree from University of Engineering and Technology, Peshawar, Pakistan, in 2009 and received Masters degree in satellite navigation from Polito di Torino, Italy, in 2012. He received his Ph.D. degree in electronics and communication engineering from Hanyang University, Seoul, South Korea in 2015. He was an Assistant Professor at the Department of Electrical Engineering, Gandhara Institute of Science and IT, Peshawar, Pakistan from August 2015 to September 2016. Dr. Saeed worked as an assistant professor at IQRA National University, Peshawar, Pakistan from October 2017 to July 2017. He is currently a Postdoctoral Research Fellow in King Abdullah University of Science and Technology (KAUST).   His current areas of interest include cognitive radio networks, underwater optical wireless communications, dimensionality reduction, and localization.
\end{IEEEbiography}

\begin{IEEEbiography}[{\includegraphics[width=1in,height=1.25in]{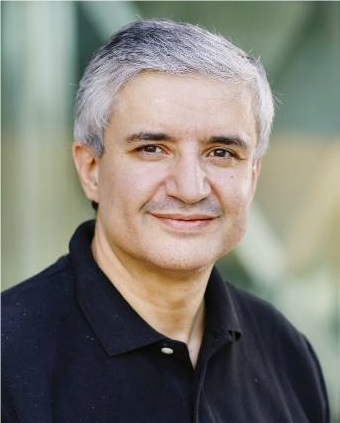}}]{Mohamed-Slim Alouini}
(S'94-M'98-SM'03-F'09)  was born in Tunis, Tunisia. He received the Ph.D. degree in Electrical Engineering
from the California Institute of Technology (Caltech), Pasadena,
CA, USA, in 1998. He served as a faculty member in the University of Minnesota,
Minneapolis, MN, USA, then in the Texas A\&M University at Qatar,
Education City, Doha, Qatar before joining King Abdullah University of
Science and Technology (KAUST), Thuwal, Makkah Province, Saudi
Arabia as a Professor of Electrical Engineering in 2009. His current
research interests include the modeling, design, and
performance analysis of wireless communication systems.
\end{IEEEbiography}

\begin{IEEEbiography}[{\includegraphics[width=1in,height=1.25in]{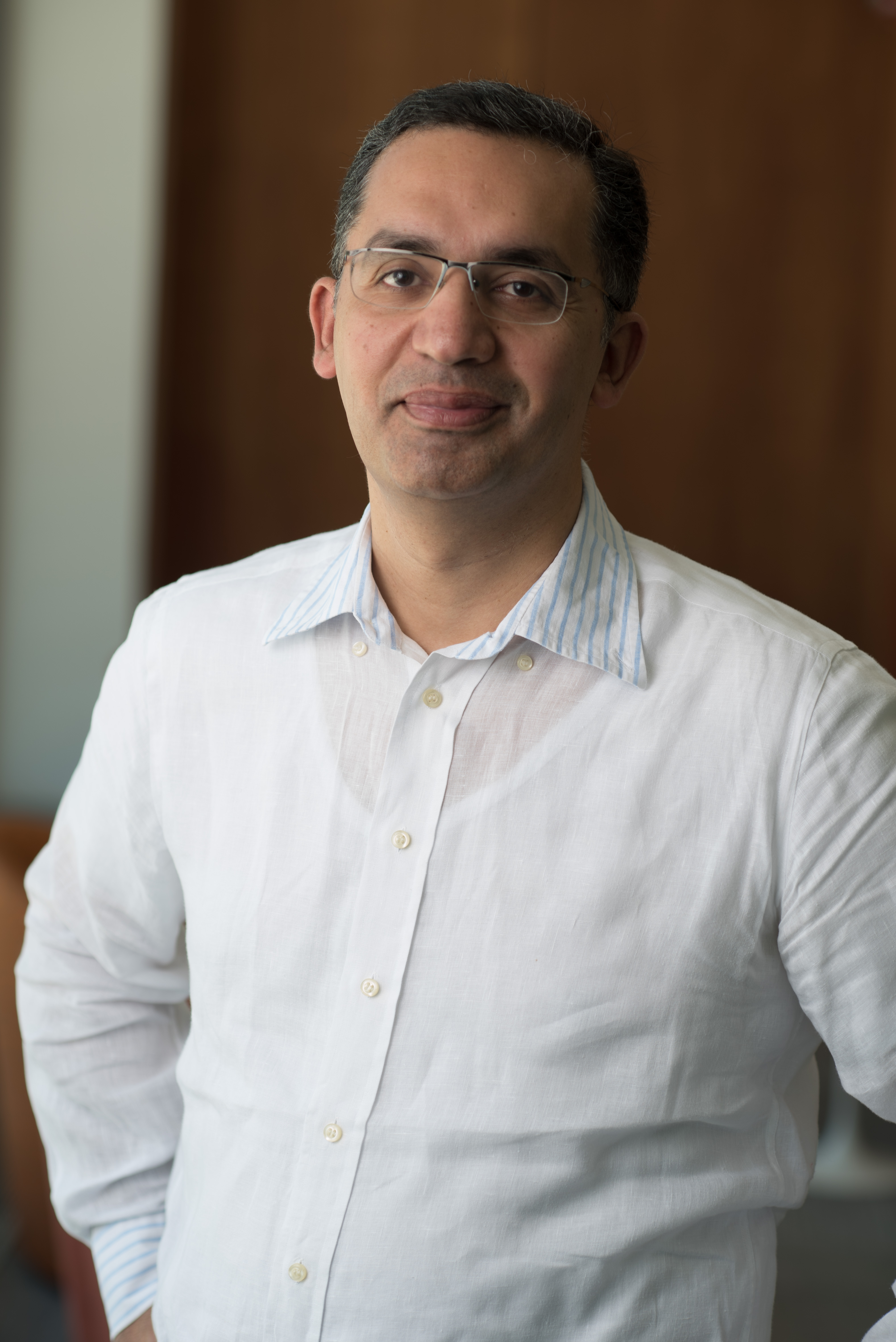}}]{Tareq Y. Al-Naffouri }
(M'10-SM'18) Tareq  Al-Naffouri  received  the  B.S.  degrees  in  mathematics  and  electrical  engineering  (with  first  honors)  from  King  Fahd  University  of  Petroleum  and  Minerals,  Dhahran,  Saudi  Arabia,  the  M.S.  degree  in  electrical  engineering  from  the  Georgia  Institute  of  Technology,  Atlanta,  in  1998,  and  the  Ph.D.  degree  in  electrical  engineering  from  Stanford  University,  Stanford,  CA,  in  2004.  
He  was  a  visiting  scholar  at  California  Institute  of  Technology,  Pasadena,  CA  in  2005  and  summer  2006.  He  was  a  Fulbright scholar  at  the  University  of  Southern  California  in  2008.  He  has  held  internship  positions  at  NEC  Research  Labs,  Tokyo,  Japan,  in  1998,  Adaptive  Systems  Lab,  University  of  California  at  Los  Angeles  in  1999,  National  Semiconductor,  Santa  Clara,  CA,  in  2001  and  2002,  and  Beceem  Communications  Santa  Clara,  CA,  in  2004.  He  is  currently  an  Associate Professor  at  the  Electrical  Engineering  Department,  King  Abdullah  University  of  Science  and  Technology  (KAUST).  His  research  interests  lie  in  the  areas  of  sparse, adaptive,  and  statistical  signal  processing  and  their  applications,  localization,  machine  learning,  and  network  information  theory.    He  has  over  240  publications  in  journal  and  conference  proceedings,  9  standard  contributions,  14  issued  patents,  and  8  pending.  Dr.  Al-Naffouri  is  the  recipient  of  the  IEEE  Education  Society  Chapter  Achievement  Award  in  2008  and  Al-Marai  Award  for  innovative  research  in  communication  in  2009.  Dr.  Al-Naffouri  has  also  been  serving  as  an  Associate  Editor  of  Transactions  on  Signal  Processing  since  August  2013. 
\end{IEEEbiography}

\end{document}